\documentclass{article}

\usepackage{arxiv}

\usepackage[utf8]{inputenc} 
\usepackage[T1]{fontenc}    
\usepackage{hyperref}       
\usepackage{url}            
\usepackage{booktabs}       
\usepackage{amsfonts}       
\usepackage{nicefrac}       
\usepackage{microtype}      
\usepackage{lipsum}
\usepackage{graphicx}
\usepackage{subcaption}
\usepackage{amsmath}
\usepackage[numbers]{natbib}
\usepackage{stmaryrd}
\usepackage{booktabs}
\usepackage{xcolor}
\usepackage{mathtools}
\usepackage{dsfont}

\graphicspath{ {./images/} }

\title{$\phi-$DeepONet: A Discontinuity Capturing Neural Operator}

\author{
 Sumanta Roy \\
  Department of Civil and Systems Engineering\\
  Johns Hopkins University \\
  Baltimore, MD 21218 \\
  \texttt{sroy41@jhu.edu} \\
   \And
 Stephen Castonguay \\
  Computational Engineering Division\\
  Lawrence Livermore National Laboratory\\
  Livermore, CA 94550 \\
  \texttt{castonguay1@llnl.gov} \\
  \And 
 Pratanu Roy \\
  Atmospheric, Earth and Energy Division\\
  Lawrence Livermore National Laboratory\\
  Livermore, CA 94550 \\
  \texttt{roy23@llnl.gov} \\
   \And 
 Michael D. Shields\thanks{Corresponding author} \\
  Department of Civil and Systems Engineering\\
  Johns Hopkins University \\
  Baltimore, MD 21218 \\
  \texttt{michael.shields@jhu.edu} \\
}

\begin{document}
\maketitle
\begin{abstract}
We present \textbf{$\boldsymbol{\phi}$-DeepONet}, a physics-informed neural operator designed to learn mappings between function spaces that may contain discontinuities or exhibit non-smooth behavior. Classical neural operators are based on the universal approximation theorem which assumes that both the operator and the functions it acts on are continuous. However, many scientific and engineering problems involve naturally discontinuous input fields as well as strong and weak discontinuities in the output fields caused by material interfaces. In $\phi$-DeepONet, discontinuities in the input are handled using multiple branch networks, while discontinuities in the output are learned through a nonlinear latent embedding of the interface. This embedding is constructed from a {\it one-hot} representation of the domain decomposition that is combined with the spatial coordinates in a modified trunk network. The outputs of the branch and trunk networks are then combined through a dot product to produce the final solution, which is trained using a physics- and interface-informed loss function. We evaluate $\phi$-DeepONet on several one- and two-dimensional benchmark problems and demonstrate that it delivers accurate and stable predictions even in the presence of strong interface-driven discontinuities.
\end{abstract}

\section{Introduction}\label{sec:introduction}

Interface problems are fundamental across science and engineering, arising wherever multiple materials, phases, or physical regimes interact. Classic examples include heat conduction in layered composites~\cite{baker1985heat,yuan2022heat}, moisture transport across materials~\cite{qin2019evaluation,roy2022multi}, reaction–diffusion systems in biological tissues~\cite{fisher2006controlling}, and flow in porous or fractured media~\cite{hosseini2021modeling}. These problems are characterized by discontinuous coefficients and/or jump conditions at interfaces, which enforce continuity of state variables and fluxes while coupling heterogeneous subdomains. Accurate and efficient resolution of such features is critical, as errors near discontinuities often dominate global solution accuracy. 

Traditionally, mesh-based numerical methods such as the finite element method (FEM), finite difference method (FDM), and finite volume methods (FVM) have been used to solve this class of problems~\cite{bramble1996finite,frei2014locally}. However, the accuracy and performance of these methods depend strongly on the problem and the chosen discretization. As a result, capturing discontinuous solution fields or sharp changes near the interface is often challenging. This difficulty increases when the geometry is irregular or non-uniform, and when the interface structure becomes complex. In such cases, generating a mesh that conforms to the interfaces becomes highly non-trivial and often turns into a major pre-processing bottleneck~\cite{valiveti2023grid,dolbow2009efficient}.

The recent rise of scientific machine learning (SciML), especially physics-informed machine learning (PIML)~\cite{karniadakis2021physics,raissi2019physics}, has helped address some of the above challenges and can work in a mesh-free manner, which greatly reduces the difficulty of mesh generation and refinement. These methods are based on the universal approximation theory of neural networks~\cite{hornik1989multilayer}, and have been successfully applied across many areas of science and engineering to capture complex and nonlinear solutions~\cite{cai2021physics,cuomo2022scientific,farea2024understanding}. However, traditional physics-informed neural networks (PINNs) fit a neural network to a single solution of a partial differential equation (PDE), which means the model must be retrained whenever any PDE parameter changes, for example material properties or initial and boundary conditions. To overcome this limitation, a class of PIML methods called neural operators~\cite{kovachki2023neural,azizzadenesheli2024neural} has been developed to learn the mapping between input and output functions defined by the governing PDE. Once the operator is learned, retraining is not required for new instances of the parametric PDE, and the model can approximate solutions drawn from the training distribution. The Deep Operator Network (DeepONet)~\cite{lu2021learning} is one of the first architectures introduced for learning such operators and is based on the universal approximation theorem for operators~\cite{chen1995universal}. In recent years, several extensions and related classes of neural operators have been proposed, including proper orthogonal decomposition DeepONets~\cite{lu2022comprehensive}, the Fourier neural operator~\cite{li2020fourier}, the wavelet neural operator (WNO)~\cite{tripura2022wavelet}, and latent DeepONets~\cite{kontolati2023learning} to name just a few.

However, the application of PIML to interface problems has been fairly limited, and several important challenges remain unaddressed. One of the early works in this direction was Extended PINNs (XPINNs)~\cite{jagtap2020extended}, which introduced domain decomposition for PINNs. Building on this idea, a number of domain-decomposition methods have been proposed specifically for interface problems, including multi-domain PINNs~\cite{zhang2022multi}, Interface-PINNs~\cite{sarma2024interface}, Adaptive Interface-PINNs~\cite{roy2025adaptive}, attention-enhanced PINNs~\cite{zheng2025ae}, heterogeneity-guided PINNs~\cite{yuan2025hg}, weight-balanced PINNs~\cite{cao2025wbpinn}, and interface-gated PINNs~\cite{zheng2025ig}. Other approaches that do not rely on domain decomposition have been introduced as well~\cite{hu2022discontinuity,hu2025discontinuity}. Notably, Hu et al.~\cite{hu2022discontinuity} introduced the idea that a discontinuous function can be represented as the restriction of a continuous function defined on a higher-dimensional space by augmenting the neural network input with a subdomain-dependent variable. This idea was later extended to include learnable categorical embeddings that encode subdomain information in a low-dimensional latent space~\cite{hu2025discontinuity}, enabling more flexible and scalable representations of piecewise-smooth functions. However, all of these methods are still PINN-based, which means they require retraining every time the parameters or boundary/initial conditions of the PDE change. More recent attempts to learn operators for interface problems include the work of Wu et al.~\cite{wu2024solving}, who proposed the interface operator network (IONet). IONet combines DeepONet with a multi-domain PINN by decomposing the physical domain into subdomains and learning separate operators in each region using different branch and trunk networks. However, as the number or complexity of interfaces grows, the number of learnable parameters increases significantly, leading to high computational cost and the same limitations found in multi-domain PINNs. Du et al.~\cite{du2024physics} introduced a Galerkin-style method where the solution is written as a linear combination of locally defined analytic basis functions tailored to the interface structure, and only the mapping from expansion coefficients to the forcing term is learned by a neural network. However, this approach is still mesh-based, which brings back many of the difficulties faced by traditional mesh-based methods for interface problems. In addition, these existing operator-learning studies focus mainly on problems with only a single interface.

To address the challenges discussed above, we build directly on the discontinuity-capturing and categorical embedding formulations of Hu et al.~\cite{hu2022discontinuity,hu2025discontinuity} and extend these ideas to the operator learning setting by introducing $\phi$-DeepONet.
In the DeepONet framework, the trunk network outputs act as basis functions whose span defines the space of representable solutions. Therefore, to accurately represent discontinuities, these features must be incorporated into the basis functions. Instead of using domain decomposition as in IONet~\cite{wu2024solving}, we embed the discontinuity directly into the outputs through a \textit{latent space} associated with the basis functions and parameterized by a modified trunk network. The latent space is constructed in two steps. First, we perform a \textit{one-hot encoding} of the sub-domains across the interface. Next, this encoding is projected into a low-dimensional \textit{latent space} that is learned during training. This latent representation acts as an additional dimension of the output function space and is provided as an extra input to the modified trunk network. Discontinuous input functions are also accommodated via the usage of multiple branch networks, following the approach proposed by Jin et al.~\cite{jin2022mionet} for handling multiple input operators. The final output functions are then approximated by taking a linear combination, through a dot product, of the outputs from the augmented trunk network and the corresponding branch networks. In this way, the proposed $\phi$-DeepONet framework incorporates interface information directly into the learned output representation, enabling the operator network to approximate discontinuous solution manifolds without relying on explicit domain decomposition.

The remainder of this paper is organized as follows. Section~\ref{sec:problem_statement} presents the governing partial differential equation for the model elliptic interface problem. Section~\ref{sec:methodology} introduces the $\phi$-DeepONet methodology, where we formulate the operator learning problem for discontinuous function spaces and discuss key implementation details. Sections~\ref{sec:numerical_examples_contint} and~\ref{sec:numerical_examples_discontint} present numerical examples that assess the performance of the proposed method. Finally, Section~\ref{sec:conclusions} summarizes the main conclusions and outlines directions for future work.

\section{Problem Statement}\label{sec:problem_statement}

We consider a class of second-order elliptic interface problems with discontinuous coefficients posed on a bounded domain $\Omega \subset \mathbb{R}^d$ with boundary $\partial \Omega$. The domain is partitioned into $P$ non-overlapping subdomains $\{\Omega_q\}_{q=1}^P$ such that
\[
\bigcup_{q=1}^P \Omega_q = \Omega,
\qquad
\Omega_p \cap \Omega_q = \emptyset \quad \text{for } p \neq q.
\]
Adjacent subdomains $\Omega_p$ and $\Omega_q$ share a sharp interface
\[
\Gamma_{pq} := \overline{\partial \Omega_p \cap \partial \Omega_q},
\]
and the collection of all interfaces is denoted by
\[
\Gamma := \bigcup_{p < q} \Gamma_{pq}.
\]

The boundary $\partial \Omega$ is equipped with an outward unit normal vector $\mathbf n_0$ and is decomposed into disjoint Dirichlet and Neumann portions. In each subdomain $\Omega_q$, we consider a general linear elliptic operator of the form
\begin{equation}
\mathcal{L}_q[s_q] := -\nabla \cdot (\mathbf K_q \nabla s_q)
+ \mathbf b_q \cdot \nabla s_q
+ c_q s_q
= f_q
\qquad \text{in } \Omega_q,
\label{eq:general_elliptic}
\end{equation}
where $\mathbf K_q$ is a symmetric positive-definite diffusion tensor, $\mathbf b_q$ and $c_q$ denote lower-order coefficients, and $f_q$ is a given source term. The coefficients and solution may be discontinuous across interfaces.

The boundary conditions on each subdomain are prescribed as
\begin{align}
\begin{split}
s_q &= \Lambda_q^{\mathrm d}
\qquad \text{on } \partial \Omega_q^{\mathrm d}, \\
\mathbf K_q \nabla s_q \cdot \mathbf n_0
&= \Lambda_q^{\mathrm n}
\qquad \text{on } \partial \Omega_q^{\mathrm n},
\end{split}
\label{eq:bc_general}
\end{align}
where $\partial \Omega_q^{\mathrm d}$ and $\partial \Omega_q^{\mathrm n}$ denote the Dirichlet and Neumann portions of the boundary, respectively, and satisfy
\[
\partial \Omega_q
=
\overline{
\partial \Omega_q^{\mathrm d}
\cup
\partial \Omega_q^{\mathrm n}
\cup
\bigcup_{p \neq q} \Gamma_{pq}
}.
\]

Across each interface $\Gamma_{pq}$, the solution satisfies the jump conditions
\begin{align}
\begin{split}
\llbracket s \rrbracket_{pq}
&= j_D^{pq}
\qquad \text{on } \Gamma_{pq}, \\
\llbracket \mathbf K \nabla s \rrbracket_{pq}
\cdot \mathbf n_{pq}
&= j_N^{pq}
\qquad \text{on } \Gamma_{pq},
\end{split}
\label{eq:jump_general}
\end{align}
where $\mathbf n_{pq}$ denotes the unit normal on $\Gamma_{pq}$ pointing outward from $\Omega_p$. The jump operator is defined as
\[
\llbracket \odot \rrbracket_{pq}
:= (\odot)_q - (\odot)_p,
\]
and $j_D^{pq}$ and $j_N^{pq}$ are prescribed scalar fields on the interface.

\begin{figure}[!hbt]
\begin{centering}
\includegraphics[width=0.3\textwidth]{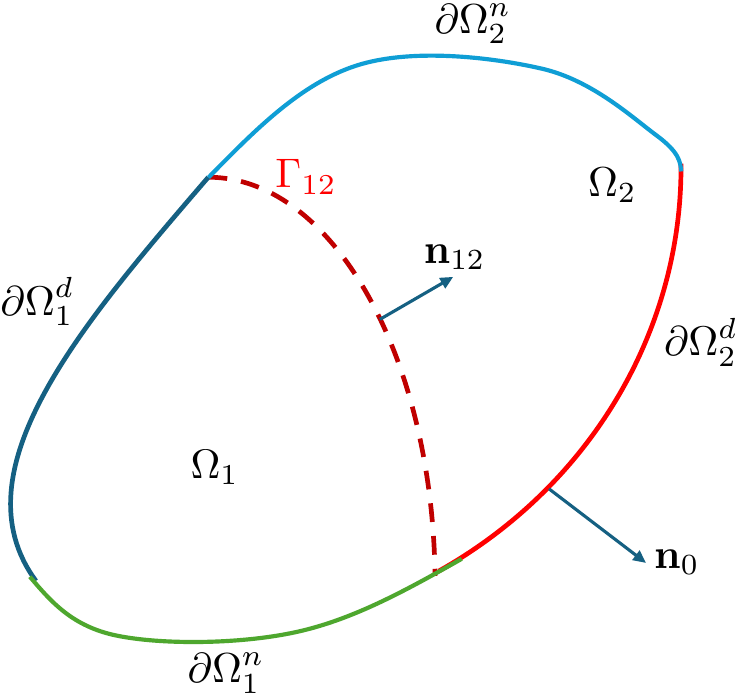}
\par\end{centering}
\caption{Schematic of the problem domain with two regions $\Omega_1$ and $\Omega_2$ separated by an interface $\Gamma_{\text{12}}$.}
\label{fig:model_problem}
\end{figure}

In this work, the numerical examples focus on Poisson-type interface problems,
\begin{equation}
\nabla \cdot (\kappa_q \nabla s_q) = -u_q
\qquad \text{in } \Omega_q,
\label{eq:eqqn_used}
\end{equation}
which correspond to the special case of Eq.~\eqref{eq:general_elliptic} with
$\mathbf b_q = \mathbf 0$, $c_q = 0$, and $\mathbf K_q = \kappa_q \mathbf I$.
This choice serves as a canonical model to clearly demonstrate the behavior of the proposed neural operator framework in the presence of sharp coefficient jumps and interface discontinuities. For clarity of exposition, Figure~\ref{fig:model_problem} shows the example of a problem domain corresponding to a single interface separating two subdomains ($P = 2$),  though the formulation and methodology extend naturally to multiple subdomains and interfaces as we will explore in the numerical examples outlined in this study.

\textbf{Physical interpretation:}  
This interface problem models situations where a physical field, such as temperature, pressure, concentration or electric potential, evolves in a heterogeneous medium with spatially varying material properties. The jump conditions prescribe how the solution and its flux behave across interfaces and are derived from fundamental physical principles such as conservation of mass or energy. In many cases, the solution and the normal component of the flux remain continuous across the interface, while the gradient of the solution exhibits directional discontinuities due to jumps in the material coefficient $\kappa$ and/or the forcing term $u_q$. In particular, for problems such as electrostatics, the tangential component of the gradient remains continuous across the interface, while the normal component may exhibit discontinuities. These features introduce sharp transitions in the solution behavior that are challenging to capture numerically and arise in applications such as heat conduction in layered materials, diffusion across membranes, and electrostatics in composite media.

\section{Methodology}\label{sec:methodology}

In this section, we first express the problem statement above in the context of operator learning version. Then we introduce the methodology and implementation strategy for our proposed methodology.

\subsection{Operator learning for discontinuous function spaces}\label{sec:methodology_operatorlearning}

We reconsider the PDE defined in Eq.~\eqref{eq:eqqn_used}. The interfaces $\Gamma$ divide the domain $\Omega$ into $P$ non-overlapping subdomains. The input function in each subdomain is denoted by $u_q$, resulting in a collection of input functions $u_1, u_2, \ldots, u_P$, which we group into a single composite input function
\[
\tilde{u} = \{u_1, u_2, \ldots, u_P\}.
\]
Similarly, the material parameters and solution fields are represented as
\[
\tilde{\kappa} = \{\kappa_1, \kappa_2, \ldots, \kappa_P\}, 
\quad
\tilde{s} = \{s_1, s_2, \ldots, s_P\},
\]
respectively. The governing PDE operator ($\mathcal{N}$), boundary condition operator ($\mathcal{B}$), and interface condition operator ($\mathcal{I}$) are collectively written as
\begin{equation}
    \mathcal{N}(\tilde{u}, \tilde{s}) = 0, \quad
    \mathcal{B}(\tilde{u}, \tilde{s}) = 0, \quad
    \mathcal{I}(\tilde{u}, \tilde{s}) = 0.
    \label{eq:PDE_representation}
\end{equation}

For each subdomain, we denote the discretized input and output functions by $\boldsymbol{u}_q$ and $\boldsymbol{s}_q$, respectively, and similarly define the vector-valued composite functions $\boldsymbol{\tilde{u}}$ and $\boldsymbol{\tilde{s}}$. We define the Banach spaces of input and output functions as
\begin{align}\label{eq:banach_spaces}
    \mathcal{U} &= \{\tilde{u} : \mathcal{X} \rightarrow \mathbb{R}^{d_u}\}, 
    \quad \mathcal{X} \subseteq \mathbb{R}^{d_x}, \\
    \mathcal{S} &= \{\tilde{s} : \mathcal{Y} \rightarrow \mathbb{R}^{d_s}\}, 
    \quad \mathcal{Y} \subseteq \mathbb{R}^{d_y}.
\end{align}

For all problems considered in this study, the forcing functions are chosen as the input functions. However, the proposed framework can be naturally extended to cases where boundary conditions, initial conditions and/or diffusion coefficients are treated as inputs. We assume that for every $\tilde{u} \in \mathcal{U}$, there exists a unique solution $\tilde{s} \in \mathcal{S}$ to the PDE defined by Eq.~\eqref{eq:eqqn_used}. This defines a ground-truth operator $\mathcal{G} : \mathcal{U} \mapsto \mathcal{S}$. Our objective is to approximate this operator using a surrogate model parameterized by $\boldsymbol{\theta}$, denoted by $\mathcal{G}_{\boldsymbol{\theta}} : \mathcal{U} \mapsto \mathcal{S}$, where $\boldsymbol{\theta} \in \mathbb{R}^{|\boldsymbol{\theta}|}$ represents the trainable parameters of the model.

The DeepONet framework~\cite{lu2021learning}, based on the universal approximation theorem for operators by Chen and Chen~\cite{chen1995universal}, employs a linear subspace approximation of the solution. Specifically, the solution is expressed as a linear combination of coefficients and basis functions. The neural network that parameterizes the coefficients is referred to as the branch network (\textit{br}), while the network that parameterizes the basis functions is called the trunk network (\textit{tr}). A direct application of DeepONet to the problem described above takes the form
\begin{equation}
    \boldsymbol{\tilde{s}}(\boldsymbol{y}) \approx 
    \mathcal{G}^{\boldsymbol{\theta}}(\boldsymbol{\tilde{u}})(\boldsymbol{y})
    \stackrel{\text{DeepONet}}{=}
    \sum_{i=1}^{K}
    br_i(\boldsymbol{\tilde{u}}; \boldsymbol{\theta})~
    tr_i(\boldsymbol{y}; \boldsymbol{\theta}),
    \label{eq:deeponet_approximation}
\end{equation}
where $\boldsymbol{\theta}$ denotes the parameters of both neural networks. The branch network takes as input a finite-dimensional representation $\boldsymbol{\tilde{u}}$ of the input function $\tilde{u}$, while the trunk network is evaluated at coordinates $\boldsymbol{y} \in \mathcal{Y}$ in the output domain. However, because the trunk network basis functions span the output function space, the induced approximation is inherently restricted to smooth functions in $\mathcal{S}$. As a result, solutions exhibiting strong or weak discontinuities are difficult to represent faithfully in practice. This limitation is evident in the numerical examples presented in Section~\ref{sec:numerical_examples_contint} and has also been reported in~\cite{wu2024solving}.

\subsection{$\phi$-DeepONet}\label{sec:methodology_phideeponet}

In this section, we introduce $\phi$-DeepONet, in which the branch and trunk networks of the classical DeepONet framework are modified to capture discontinuities in both the input and output functions. Inspired by the multi-input operator framework~\cite{jin2022mionet}, we first decompose the standard branch network into multiple smaller branch networks, denoted by $br^q$, each corresponding to a subdomain $q$. Each of these domain-specific branch networks learns the input function $\boldsymbol{u}_q$ defined on the corresponding subdomain of the input function space $\mathcal{U}$. To model discontinuities in the output function space $\mathcal{S}$, information about the interfaces is embedded through an additional latent variable $\boldsymbol{\phi}$. This latent variable is passed to a modified trunk network together with the output coordinates $\boldsymbol{y}$. The resulting $\phi$-DeepONet approximation is given by
\begin{align}
    \boldsymbol{\tilde{s}}^{\boldsymbol{\theta}}(\boldsymbol{y}) 
    &\approx \mathcal{G}^{\boldsymbol{\theta}}(\boldsymbol{\tilde{u}})(\boldsymbol{y})
    \stackrel{\phi\text{-DeepONet}}{=} 
    \sum_{i=1}^{K}
    br^1_i(\boldsymbol{u}_1;\boldsymbol{\theta})
    br^2_i(\boldsymbol{u}_2;\boldsymbol{\theta})
    \cdots
    br^P_i(\boldsymbol{u}_P;\boldsymbol{\theta})
    \, tr_i(\boldsymbol{y},\boldsymbol{\phi};\boldsymbol{\theta}), \\
    &=
    \sum_{i=1}^{K}
    \left(
    \prod_{q=1}^{P}
    br^q_i(\boldsymbol{u}_q;\boldsymbol{\theta})
    \right)
    tr_i\!\left(\boldsymbol{y},\boldsymbol{\phi}(\boldsymbol{y});\boldsymbol{\theta}\right).
    \label{eq:phideeponet_approximation}
\end{align}
We now describe the modified branch and trunk networks in more detail.

\medskip
\noindent
\textbf{Modified branch network.}  
For the model problem described above, where the interfaces partition the domain $\Omega$ into $P$ subdomains, we consider $N$ discontinuous input functions
\[
\left(\tilde{u}^{(1)}, \tilde{u}^{(2)}, \ldots, \tilde{u}^{(N)}\right),
\]
where each input function $\tilde{u}^{(i)}$ is decomposed into $P$ piecewise continuous components,
\[
\left(u_1^{(i)}, u_2^{(i)}, \ldots, u_P^{(i)}\right).
\]

These $N$ input functions are available in finite-dimensional form as
\[
\left(\boldsymbol{\tilde{u}}^{(1)}, \boldsymbol{\tilde{u}}^{(2)}, \ldots, \boldsymbol{\tilde{u}}^{(N)}\right),
\]
corresponding to function values sampled at fixed sensor locations. Note that the number of sensors may differ across subdomains. For the $q$-th subdomain, the branch network $br^q(\boldsymbol{\tilde{u}}_q^{(i)})$ takes as input the vector $\boldsymbol{\tilde{u}}_q^{(i)}$, which contains samples of the function $u^{(i)}$ restricted to that subdomain. As a result, a problem with $P$ subdomains uses $P$ branch networks,
\[
\left(br^1, br^2, \ldots, br^p\right).
\]
The outputs of these branch networks are multiplied and combined with the modified trunk network to form the solution approximation in Eq.~\eqref{eq:phideeponet_approximation}.

\medskip
\noindent
\textbf{Modified trunk network.}  
To capture discontinuities in the solution space $\mathcal{S}$, the trunk network must also be modified. Unlike IONet~\cite{wu2024solving}, which introduces a separate trunk network for each subdomain, we embed interface information within a single trunk network using an additional latent variable. This approach is motivated by the observation that a $d$-dimensional discontinuous function can be represented as a smooth function in $(d+1)$ higher-dimensional space~\cite{hu2022discontinuity}. In this setting, the interfaces act as zero-level sets of an auxiliary embedding.

Accordingly, we introduce an \textit{augmented trunk network} that takes both the spatial coordinates $\boldsymbol{y}$ and the latent variable $\boldsymbol{\phi}$ as inputs, written as $tr(\boldsymbol{y}, \boldsymbol{\phi})$ (the dependence on parameters $\boldsymbol{\theta}$ is omitted here for clarity). The latent mapping $\boldsymbol{\phi} : \mathbb{R}^d \rightarrow \mathbb{R}^D$ assigns each spatial location $\boldsymbol{y} \in \mathbb{R}^d$ to its corresponding subdomain representation, where $d$ is the spatial dimension and $D$ is the latent dimension. Following the ideas introduced in~\cite{hu2025discontinuity}, we consider three different strategies for constructing this latent embedding.

\begin{enumerate}
    \item \textbf{Scalar embedding (SE):}  
    As the name suggests, we first consider a scalar-valued latent embedding
    \[
    \boldsymbol{\phi} : \mathbb{R}^d \rightarrow \mathbb{R},
    \qquad (D = 1).
    \]
    Such an embedding can be constructed in several ways, for example by extracting dominant modes of the solution space or by parameterizing $\boldsymbol{\phi}$ using a neural network. In this work, however, we adopt the simplest possible approach by prescribing constant scalar values within each subdomain. Specifically, we define
    \[
    \boldsymbol{\phi}(\boldsymbol{y}) = \gamma_q,
    \qquad \boldsymbol{y} \in \Omega_q,
    \]
    where $\gamma_q \in \mathbb{I}$ is a scalar associated with the $q$-th subdomain. We choose
    \[
    \gamma_q = q, \qquad q = 1, \dots, P,
    \]
    so that the latent variable directly encodes the subdomain index. While this approach provides a simple way to include subdomain information, it can become limiting as the number of interfaces increases or their geometry becomes more complex. Moreover, in many machine learning settings, prescribing such constants \emph{a priori} is impractical and lacks flexibility. These limitations motivate the use of learnable categorical embeddings, which are described next.

    \item \textbf{Categorical embedding (CE):}  
    We also use a categorical embedding in which the latent variable $\boldsymbol{\phi}$ is learned during training. This approach is particularly useful when explicit labels or ordering of subdomains are not available. We define a \textit{one-hot} indicator mapping
    \[
    \mathds{1} : \mathbb{R}^d \rightarrow \mathbb{R}^P,
    \qquad
    \]
    where the $q$-th component of $\mathds{1}(\boldsymbol{y})$ is given by
    \[
    \big[\mathds{1}(\boldsymbol{y})\big]_q =
    \begin{cases}
    1, & \text{if } \boldsymbol{y} \in \Omega_q, \\
    0, & \text{otherwise},
    \end{cases}
    \qquad q = 1, \dots, P.
    \]
    Here, $\{\Omega_q\}_{q=1}^{{P}}$ forms a partition of the computational domain, ensuring that $\mathds{1}(\boldsymbol{y})$ is a \textit{one-hot} vector with exactly one nonzero entry. This construction avoids imposing any artificial ordering on the subdomains. The goal is to learn a latent representation $\boldsymbol{\phi}$ from the categorical information encoded in $\mathds{1}(\boldsymbol{y})$. To achieve this, we introduce an embedding matrix $\mathbf{E} \in \mathbb{R}^{D \times P}$ and define
    \begin{equation}
        \boldsymbol{\phi}(\boldsymbol{y}) = \mathbf{E}\,\mathds{1}(\boldsymbol{y}),
        \label{eq:categorical_embedding}
    \end{equation}
    where $\boldsymbol{\phi} : \mathbb{R}^d \rightarrow \mathbb{R}^D$. The embedding matrix $\mathbf{E}$ is learned jointly with the network parameters. Compared to the SE approach, this method offers greater flexibility, since the latent representation is inferred automatically from data rather than being prescribed \emph{a priori}. In particular, this allows the model to adaptively learn representations that are consistent with the underlying interface structure, rather than relying on fixed scalar labels tied to subdomain indices. As a result, the embedding can more effectively capture complex interface configurations, especially in cases where the geometry or number of subdomains is not known in advance or is difficult to parameterize explicitly. As noted in~\cite{hu2025discontinuity}, this categorical embedding is closely related to entity embedding techniques~\cite{guo2016entity}, which map high-cardinality categorical variables to low-dimensional continuous spaces.

    \item \textbf{Non-linear categorical embedding (Non-linear CE):}  
    The categorical embedding defined in Eq.~\eqref{eq:categorical_embedding} corresponds to a linear projection of the \textit{one-hot} representation into the latent space (similar to traditional methods like Principal Component Analysis or Proper orthogonal decomposition). To further increase the expressive power of the embedding, we introduce a non-linear transformation. Specifically, we apply an element-wise activation function
    \[
    \sigma : \mathbb{R}^D \rightarrow \mathbb{R}^D,
    \]
    leading to
    \begin{equation}
        \boldsymbol{\phi}(\boldsymbol{y}) = \sigma\!\left( \mathbf{E}\,\mathds{1}(\boldsymbol{y}) \right),
        \label{eq:categorical_embedding_nonlinear}
    \end{equation}
    where $\sigma(\cdot)$ may be chosen as a standard activation function such as ReLU, $\tanh$, or GELU. This non-linear embedding allows the model to learn more complex latent representations of the interface structure. We emphasize that this formulation represents a minimal extension to introduce non-linearity. For problems exhibiting higher complexity, more expressive non-linear dimensionality reduction techniques such as autoencoders may be employed.
\end{enumerate}

\begin{figure}[!hbt]
\begin{centering}
\includegraphics[width=1.0\textwidth]{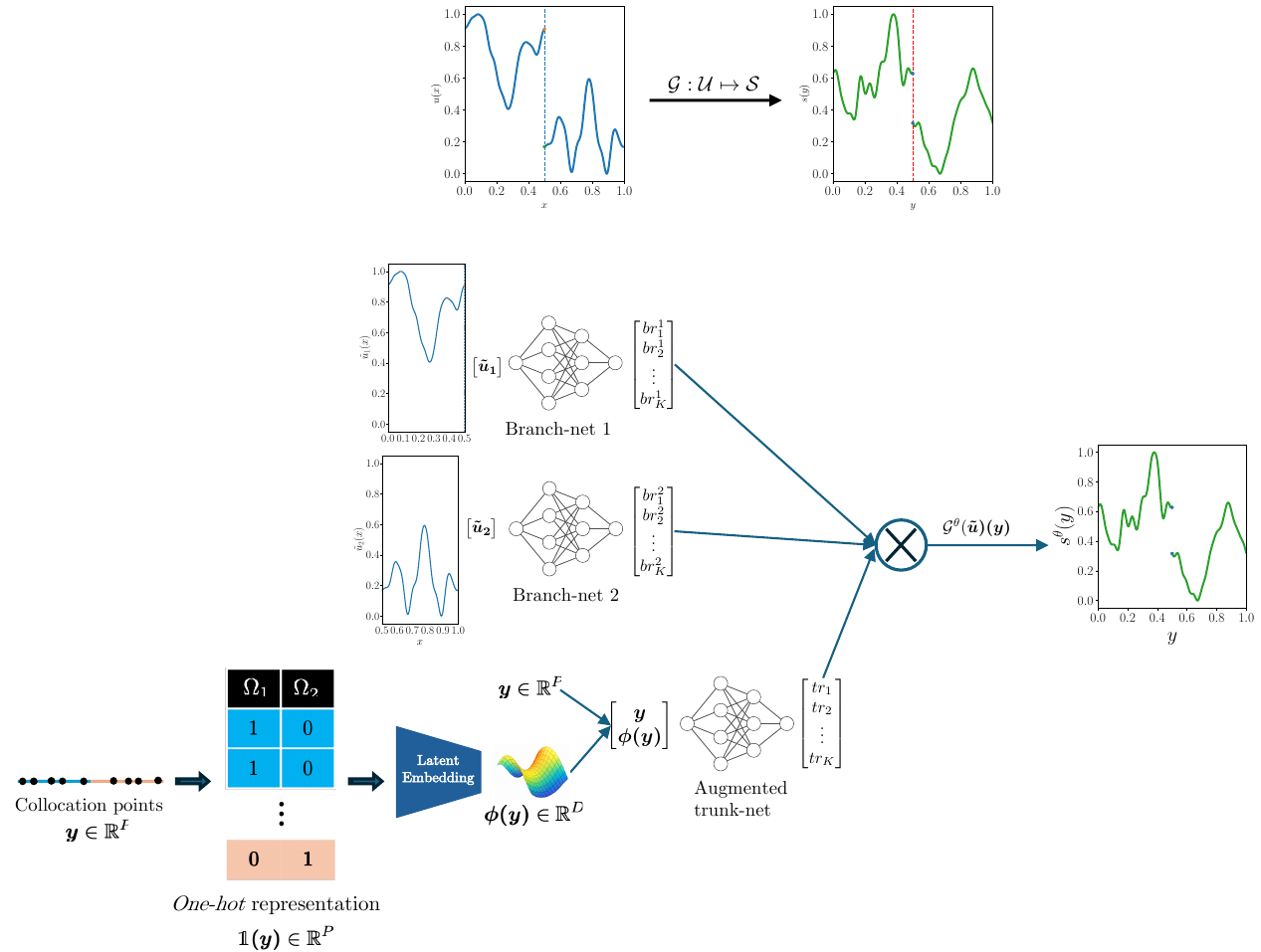}
\par\end{centering}
\caption{Schematic of the architecture of $\phi-$DeepONet for approximating the operator $\mathcal{G}:u \mapsto s$.}\label{fig:phi_deeponet_schematic}
\end{figure}

Figure~\ref{fig:phi_deeponet_schematic} shows a schematic of the $\phi$-DeepONet framework. The architecture is consistent, regardless of embedding (SE, CE, or Non-linear CE).
We denote the combined parameters of all neural networks (along with the entries of embedding matrix $\mathbf{E}$) by $\boldsymbol{\theta}$. These parameters are trained using a physics- and interface-informed loss function, defined as
\begin{multline}
    \mathcal{L}(\boldsymbol{\theta}) 
    = \mathcal{L}_{\text{PDE}} + \mathcal{L}_{\text{BC}} + \mathcal{L}_{\text{I}}
    =\\ \frac{1}{Nm} \sum_{i=1}^{N} \sum_{j=1}^{m} 
    \left|
    \mathcal{N}\!\left( \boldsymbol{\tilde{u}}^{(i)}(\boldsymbol{y}_j),
    \boldsymbol{\tilde{s}}^{\boldsymbol{\theta}}(\boldsymbol{y}_j) \right)
    \right|^2
    +
    \frac{1}{Nb} \sum_{i=1}^{N} \sum_{j=1}^{b} 
    \left|
    \mathcal{B}\!\left( \boldsymbol{\tilde{u}}^{(i)}(\boldsymbol{y}_j),
    \boldsymbol{\tilde{s}}^{\boldsymbol{\theta}}(\boldsymbol{y}_j) \right)
    \right|^2
    +
    \frac{1}{Nt} \sum_{i=1}^{N} \sum_{j=1}^{t} 
    \left|
    \mathcal{I}\!\left( \boldsymbol{\tilde{u}}^{(i)}(\boldsymbol{y}_j),
    \boldsymbol{\tilde{s}}^{\boldsymbol{\theta}}(\boldsymbol{y}_j) \right)
    \right|^2,
    \label{eq:loss_function}
\end{multline}
where $\mathcal{N}$, $\mathcal{B}$, and $\mathcal{I}$ denote the PDE, boundary, and interface condition operators, respectively. Here, $N$ is the number of input function realizations, $m$ is the number of collocation points used to evaluate the PDE residual, and $b$ and $t$ are the numbers of points used to enforce the boundary and interface conditions.

\medskip
\noindent
\textbf{Hard enforcement of constraints:}  
The loss function in Eq.~\eqref{eq:loss_function} uses a soft-constraint formulation, in which boundary and interface conditions are enforced through penalty terms. As an alternative, constraints may be imposed by reparameterizing the neural operator approximation so that the constraints are satisfied by construction. Specifically, we write
\begin{equation}
\boldsymbol{\tilde{s}}_{\boldsymbol{\theta}}(\boldsymbol{y})
=
\lambda_s(\boldsymbol{y}) \,
\boldsymbol{\tilde{s}}^{\boldsymbol{\theta}}(\boldsymbol{\tilde{u}}, \boldsymbol{y}, \boldsymbol{\phi})
+ H(\boldsymbol{y}),
\qquad \boldsymbol{y} \in \Omega,
\label{eq:hard_constraint}
\end{equation}

where $H(\boldsymbol{y})$ is a known, problem-dependent function chosen to exactly satisfy the prescribed Dirichlet boundary conditions on the external boundary $\partial \Omega$. The function $H(\boldsymbol{y})$ depends only on the geometry and boundary data and is independent of the neural network parameters. The function $\lambda_s(\boldsymbol{y})$ is a distance-type function that vanishes on the constrained boundary locations and remains strictly positive in the interior of the domain. For example, in a rectangular domain, $\lambda_s$ may be constructed as the product of distances to the Dirichlet boundaries. More general constructions of distance functions, including those suitable for complex geometries and mixed boundary conditions, can be obtained using R-functions and mean value potential theory~\cite{sukumar2022exact}.

In all numerical experiments presented in this work, only Dirichlet boundary conditions on $\partial \Omega$ are enforced using this hard-constraint formulation. While similar reparameterizations may also be employed to hard-enforce interface conditions on internal boundaries $\Gamma_{pq}$, such extensions are not considered here. Readers interested in hard enforcement of both essential and mixed boundary or interface conditions are referred to the work of Sukumar et al.~\cite{sukumar2022exact}.

\section{Numerical Examples with Continuous Inputs}\label{sec:numerical_examples_contint}

We demonstrate the capability of $\phi$-DeepONet to solve interface problems through several example problems. We divide the problems into two types based on how they are constructed. In the first type, the input functions fed to the neural operator are continuous, but due to the discontinuity in the material parameters, the resulting output function gradients become discontinuous. In the second type, both the inputs and the output gradients (as well as the material parameters) contain discontinuities. In this section, we focus on problems of the first type. All input functions $u(\mathbf{x})$ are generated as samples from Gaussian random fields (GRFs) with mean $\mu$ and length scale $l_s$:
\begin{equation}
u \sim \mathcal{GP}\big(\mu,\, k_{l_s}(\mathbf{x}_1, \mathbf{x}_2)\big),\label{eq:gp_inputs_a}
\end{equation}
where the covariance function is the squared-exponential (exponential quadratic) kernel
\begin{equation}
k_{l_s}(\mathbf{x}_1,\mathbf{x}_2)
= \exp\!\left(
    -\frac{\|\mathbf{x}_1 - \mathbf{x}_2\|^2}{2\, l_s^2}
  \right).\label{eq:gp_inputs_b}
\end{equation}
The entire dataset is divided into training ($M_\text{train}$) and test ($M_\text{test}$) sets, and the collocation points for each are chosen randomly over the domain. Model performance is evaluated using the relative $L_2$ error metric, defined as $||\boldsymbol{s_\theta} - \boldsymbol{s}||_2/||\boldsymbol{s}||_2$, where $\|\cdot\|$ denotes the $L_2$ norm. Unless otherwise noted, all models are trained using the \textit{Adam}~\cite{kingma2017adam} and \textit{SOAP} optimizers~\cite{vyas2024soap} with their default settings, a learning rate of $5\times10^{-3}$, and a maximum of 5000 training epochs. The parameters (weights and biases) of the branch and trunk networks are initialized using the Xavier initialization scheme~\cite{glorot2010understanding}. All models are implemented in the JAX machine learning framework~\cite{sapunov2024deep} and trained on an Apple M3 Max CPU. To compare the computational cost of different models, we define a relative cost metric, denoted as \textit{Cost}, as the ratio of the training time of a given model with respect to a reference model. For instance, if $\text{Cost} = 1$ for Model A and $\text{Cost} = 2.5$ for Model B, this indicates that Model B requires 2.5 times the training time of Model A.

\subsection{One-dimensional problem with single interface}\label{sec:1d_oneint}

As a first example, we consider a one-dimensional problem defined on the domain $\Omega=[0,1]$ with the presence of an interface $\zeta_\text{int}=0.5$ thus dividing the sub-domain into $\Omega_1=[0,\zeta_\text{int}]$ and $\Omega_2=[\zeta_\text{int},1]$ such that $\Omega=\Omega_1 \cup \Omega_2$. We seek to train the neural operator to satisfy the PDE given by:
\begin{align}
  \begin{split}
      \frac{d}{dy}\left({\kappa_q} \frac{d{s_q}}{dy}\right)&= u_q \quad \text{in} ~ \Omega_q, \\
       s_{1} &= 0  \quad \text{at}~y=0, \\
       s_{2} &= 0  \quad \text{at}~y=1, \\
       \llbracket s \rrbracket &=0, ~\text{at}~ y={0.5}, \\
       \left \llbracket {\kappa\frac{ds}{dy}}  \right \rrbracket &=0 ~\text{at}~ y={0.5}. 
  \end{split}
\end{align}\label{eq:1d_oneint}

\begin{figure}[!hbt]
    \centering
    \begin{subfigure}[b]{0.265\textwidth}
        \centering
        \includegraphics[width=\textwidth]{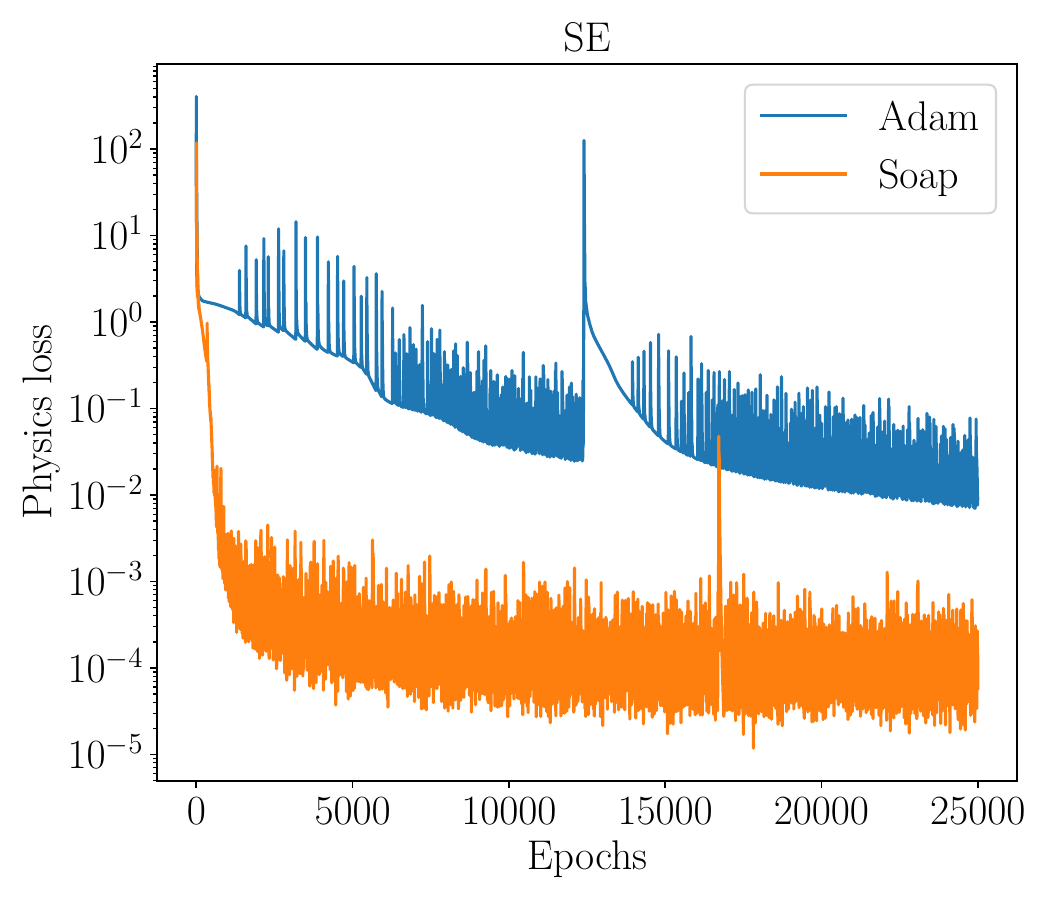}
        \caption{}
        \label{fig:1d_one_int_phiDON_SE_convergence}
    \end{subfigure}
    \hspace{2.0pt}
    \begin{subfigure}[b]{0.23\textwidth}
        \centering
        \includegraphics[width=\textwidth]{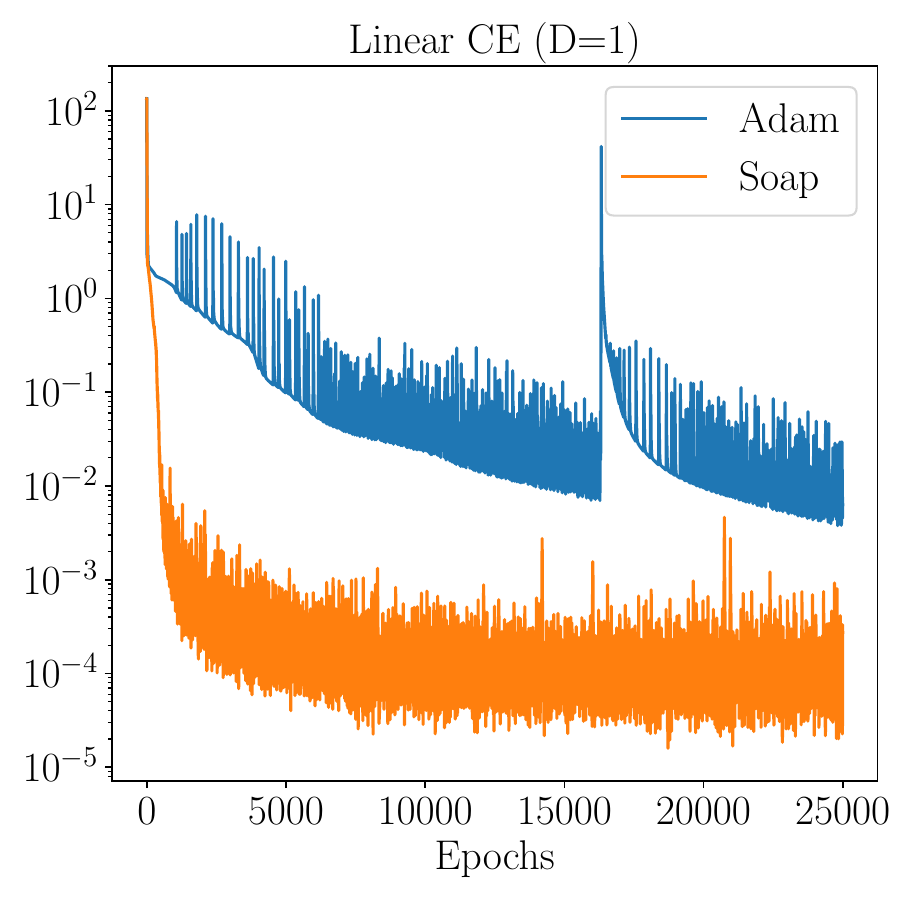}
        \caption{}
        \label{fig:1d_one_int_phiDON_CELinear_convergence}
    \end{subfigure} 
    \hspace{2.0pt}
    \begin{subfigure}[b]{0.23\textwidth}
        \centering
        \includegraphics[width=\textwidth]{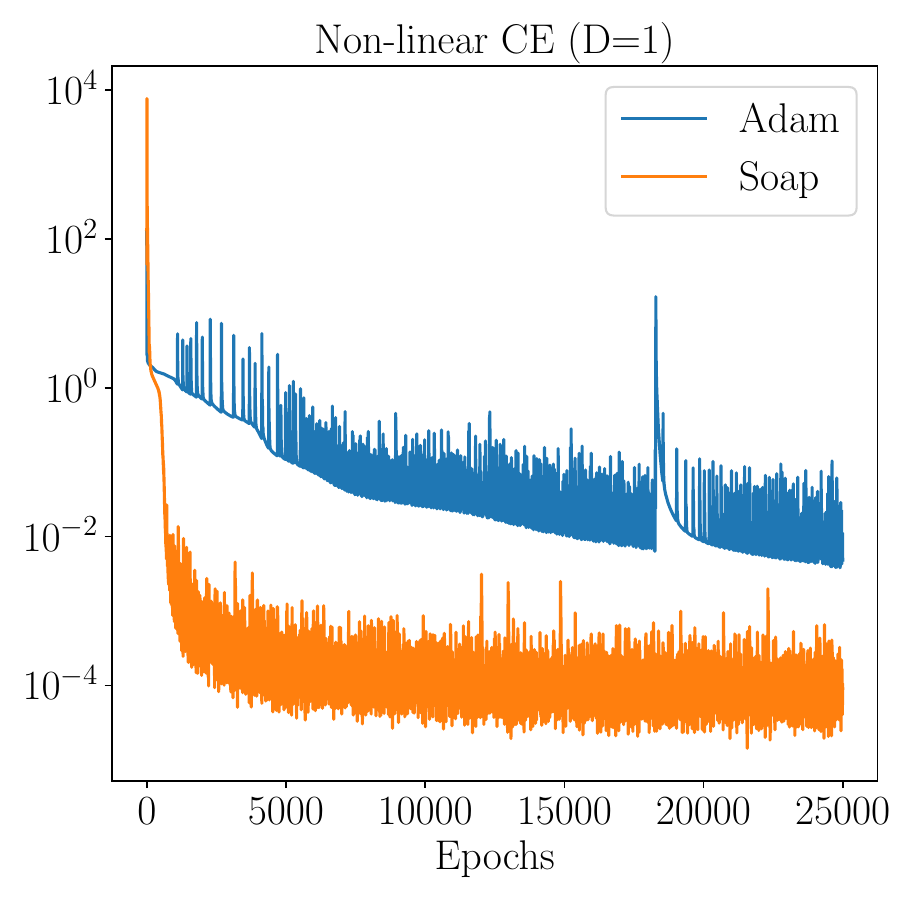}
        \caption{}
        \label{fig:1d_one_int_phiDON_CEnonLinear_D=1_convergence}
    \end{subfigure}
    \hspace{2.0pt}
    \begin{subfigure}[b]{0.23\textwidth}
        \centering
        \includegraphics[width=\textwidth]{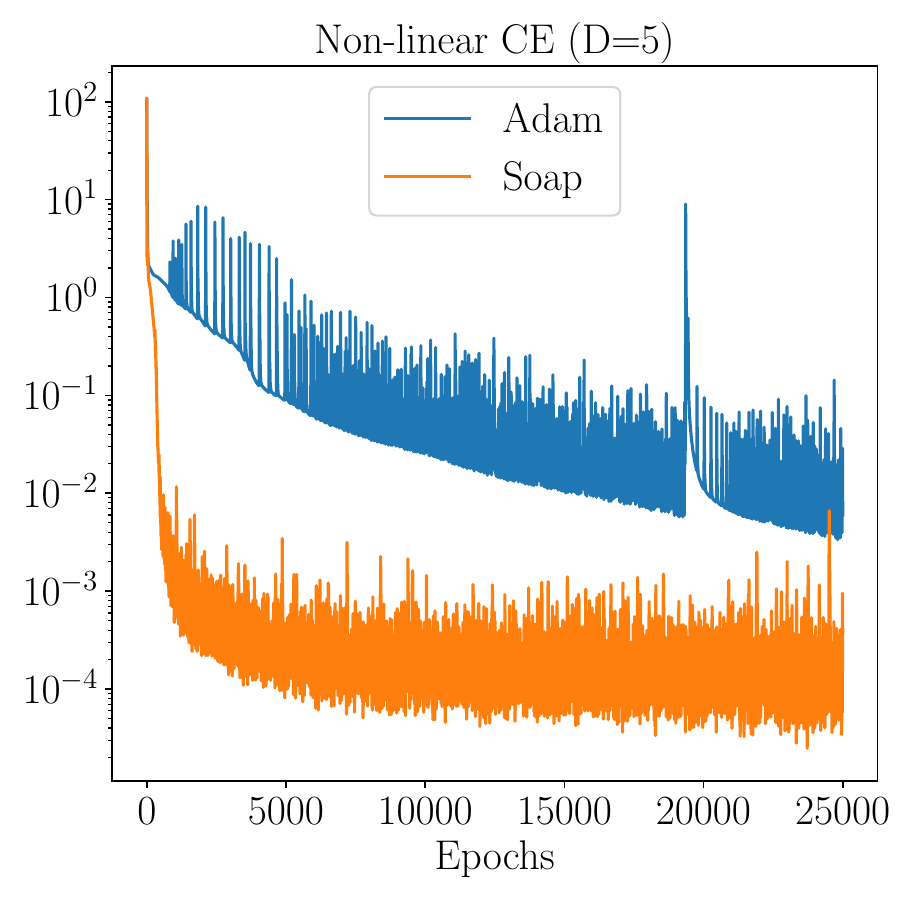}
        \caption{}
        \label{fig:1d_one_int_phiDON_CEnonLinear_D=5_convergence}
    \end{subfigure}\\
    \caption{Convergence profiles (physics loss or MSE vs epochs) for the various $\phi-$DeepONet models using two optimizers: \textit{Adam} and \textit{SOAP}. SE denote the scalar embedding, CE denotes the categorical embedding, and non-linear CE denotes the categorical embedding matrix wrapped with a non-linear ($tanh$) activation function. The latent embedding dimension is denoted by $D$.}\label{fig:1d_one_int_phiDON_convergence}
\end{figure}

Our goal is to approximate the operator $G: u(x) \rightarrow s(y)$, where $u(x)$ is modeled as a GRF with mean $\mu = 1.0$ and length scale $l_s = 0.2$. The material constants for the two regions are set as $\kappa_1 = 5$ and $\kappa_2 = 0.1$. As noted earlier, the input functions $u(x)$ are continuous at the interface ($u_1 = u_2$), but the derivative of the output function becomes discontinuous because of the jump in material properties. We use 1000 training samples and evaluate performance on 250 test samples. We compare several variants of $\phi$-DeepONet on this problem, including the scalar embedding (SE) and categorical embedding (CE) frameworks. For the CE models, we test both linear embeddings and non-linear embeddings, where the embedding matrix is passed through a \textit{tanh} activation function. Figure~\ref{fig:1d_one_int_phiDON_convergence} shows the convergence curves for all these models using both optimizers. The figure also includes results for non-linear CE models with different embedding dimensions ($D=1$ and $D=5$). Across all models, the \textit{SOAP} optimizer consistently gives the best convergence, independent of the embedding dimension. This can be attributed to the fact that \textit{SOAP} is a second-order, pre-conditioned optimizer that approximates curvature by cheaply estimating the diagonal of the Hessian, whereas \textit{Adam} is only a first-order method. As a result, \textit{SOAP} takes larger steps in flatter directions and smaller steps in steeper ones, allowing it to reach the minimum faster and more stably. Therefore, for all remaining experiments, we use the \textit{SOAP} optimizer.

Table~\ref{tab:1d_oneint_errors} reports the mean relative $L_2$ errors on the test dataset for the different $\phi$-DeepONet variants. The ground truth solutions were obtained using a finite difference method with harmonic averaging of the coefficients. We also compare the $\phi$-DeepONet results with the standard physics-informed DeepONet~\cite{wang2021learning} and the physics-informed IONet~\cite{wu2024solving} frameworks. As seen in the table, all $\phi$-DeepONet models clearly outperform the standard DeepONet, with mean relative test errors of $\mathcal{O}(10^{-2})$-$\mathcal{O}(10^{-3})$, while the latter only achieves errors of $\mathcal{O}(10^{-1})$. Among the $\phi$-DeepONet variants, the non-linear categorical models achieve the best accuracy, outperforming both the linear categorical model and the scalar embedding model. Among the non-linear models, the effect of the latent dimension is noticeable only for small values of $D$. Once $D \ge 3$, this effect becomes minimal. Compared to the physics-informed IONet, the $\phi$-DeepONet achieves a similar level of approximation accuracy (i.e., the same order of magnitude in error). However, IONet incurs a slighty higher computational cost, with $\textit{Cost} = 1.26$, whereas all $\phi$-DeepONet variants exhibit comparable training times. Although this might not be that significant for a toy problem with a single interface, this scales up when the number and spread of interfaces increase, as we will see in the next few examples.

\begin{table}[h!]
\centering
\begin{tabular}{l c c c c c c c c c}
\toprule
& \textbf{SE} 
& \textbf{CE} 
& \multicolumn{5}{c}{\textbf{Non-linear CE}} 
& \textbf{DeepONet}
& \textbf{IONet} \\
\cmidrule(lr){4-8}
&  &  & $D=1$ & $D=2$ & $D=3$ & $D=4$ & $D=5$ &  \\
\midrule
\textbf{Rel.~$L_2$ error}
& $5.69\text{e-}2$
& $2.27\text{e-}2$
& $9.44\text{e-}3$
& $8.21\text{e-}3$
& $6.84\text{e-}3$
& $3.73\text{e-}3$
& $3.45\text{e-}3$
& $7.22\text{e-}1$
& $1.31\text{e-}3$ \\
\textbf{Cost}
& 1.00
& 1.01
& 1.02
& 1.02
& 1.02
& 1.01
& 1.02
& 1.13
& 1.26 \\
\bottomrule
\end{tabular}
\caption{Relative $L_2$ errors on the test set for the various $\phi$-DeepONet frameworks (along with the standard physics-informed DeepONet~\cite{wang2021learning} and physics-informed IONet models~\cite{wu2024solving}) for the 1D problem with one interface (Section~\ref{sec:1d_oneint}).}\label{tab:1d_oneint_errors}
\end{table}

\begin{figure}[!hbt]
    \centering
    \begin{subfigure}[b]{0.77\textwidth}
        \centering
        \includegraphics[width=\textwidth]{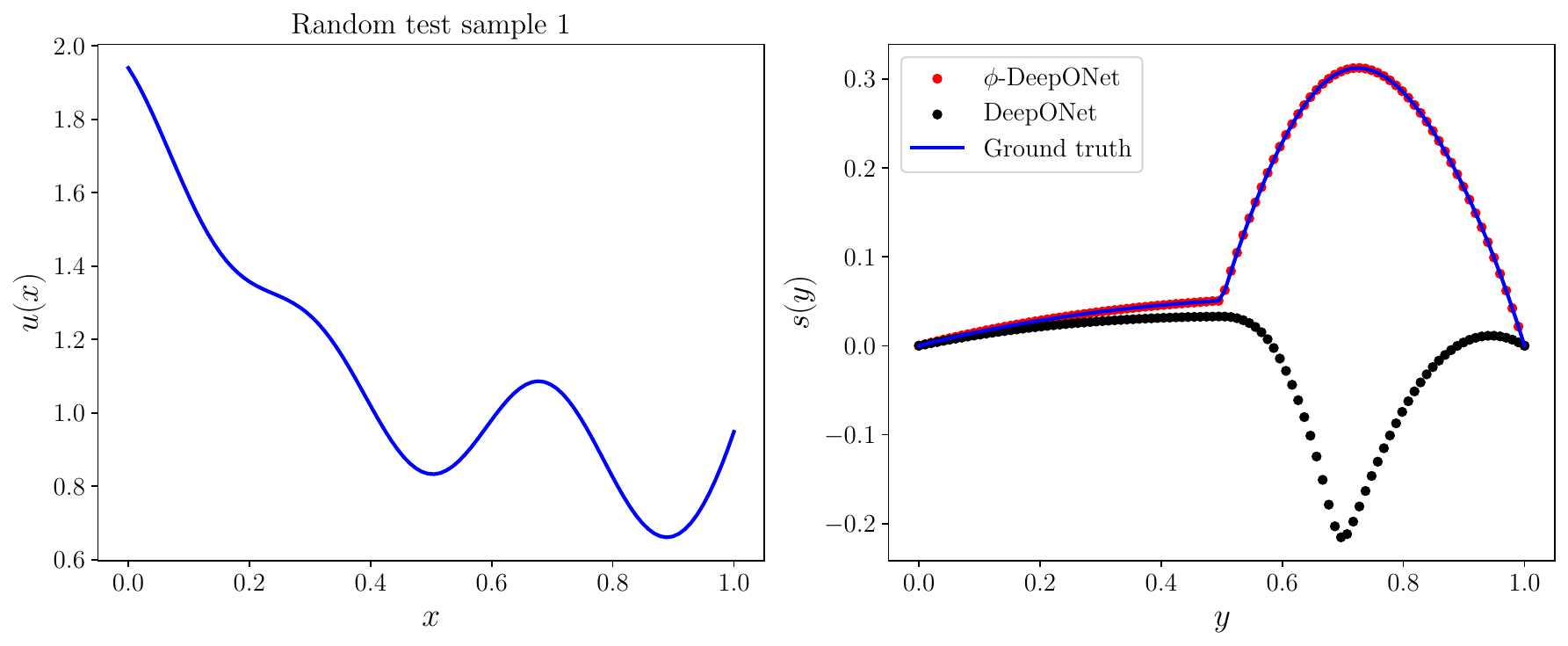}
        \caption{}
        \label{fig:1d_oneint_compare_test1}
    \end{subfigure}\\
    \centering
    \begin{subfigure}[b]{0.77\textwidth}
        \centering
        \includegraphics[width=\textwidth]{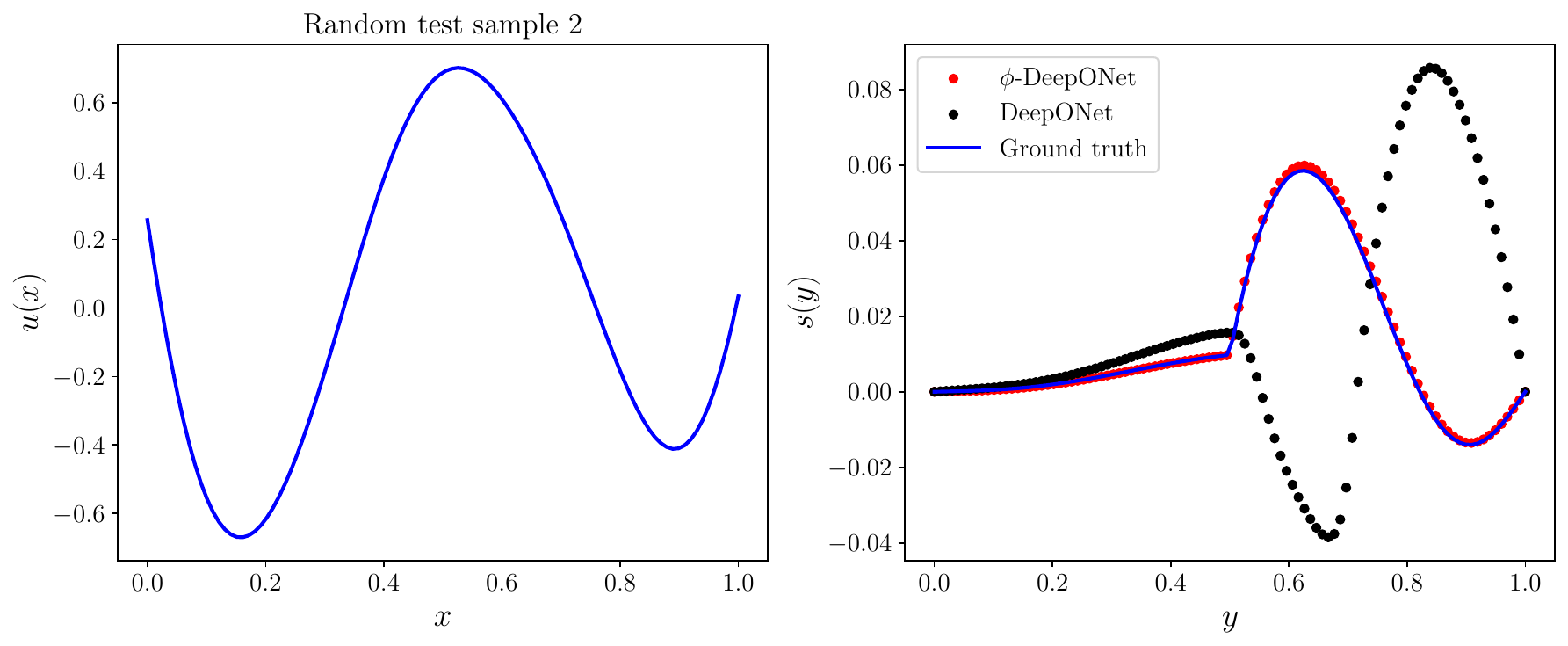}
        \caption{}
        \label{fig:1d_oneint_compare_test2}
    \end{subfigure}\\
    \centering
    \begin{subfigure}[b]{0.4\textwidth}
        \centering
        \includegraphics[width=\textwidth]{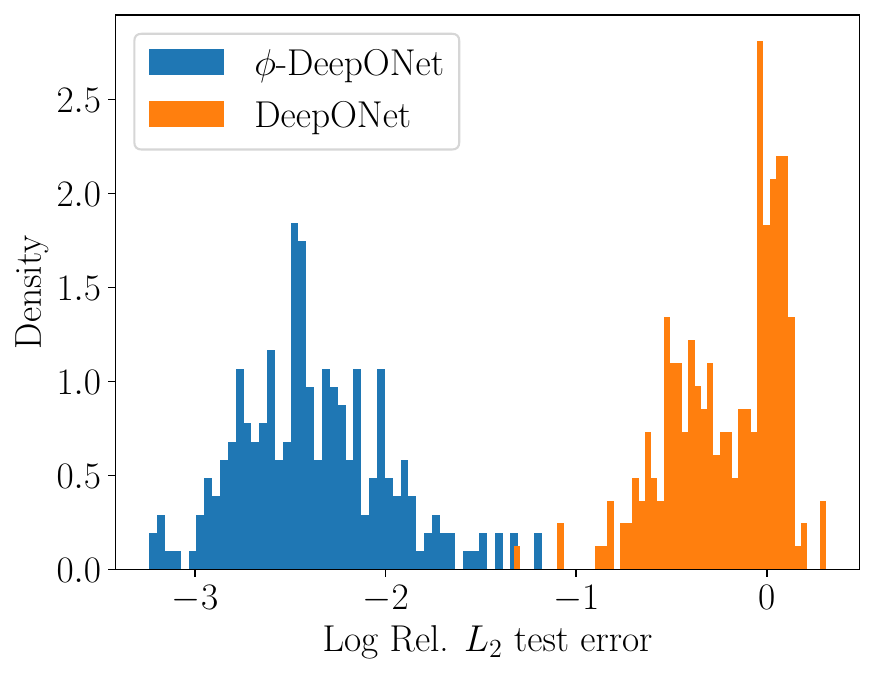}
        \caption{}
        \label{fig:comparison_1doneint_errorplot}
    \end{subfigure}\\
    \caption{1D single-interface problem: performance of the $\phi$-DeepONet framework (and the standard DeepONet) on two random test samples (subfigures (a) and (b)), along with the distribution of the test errors for both models (subfigure (c)).}
    \label{fig:1d_oneint_examples_with_convergence}
\end{figure}

In Figure~\ref{fig:1d_oneint_examples_with_convergence}, we show two random examples from the test set and compare the predictions of the proposed $\phi$-DeepONet model with those of the standard DeepONet. We observe that the standard DeepONet fails to capture the correct solution, while $\phi$-DeepONet predicts it accurately. Figure~\ref{fig:comparison_1doneint_errorplot} presents the distribution of test errors for both frameworks, where $\phi$-DeepONet clearly achieves lower errors across the entire test dataset. The ground-truth solutions are obtained using a finite difference method based on an immersed interface method.

\begin{figure}[!hbt]
\begin{centering}
\includegraphics[width=0.5\textwidth]{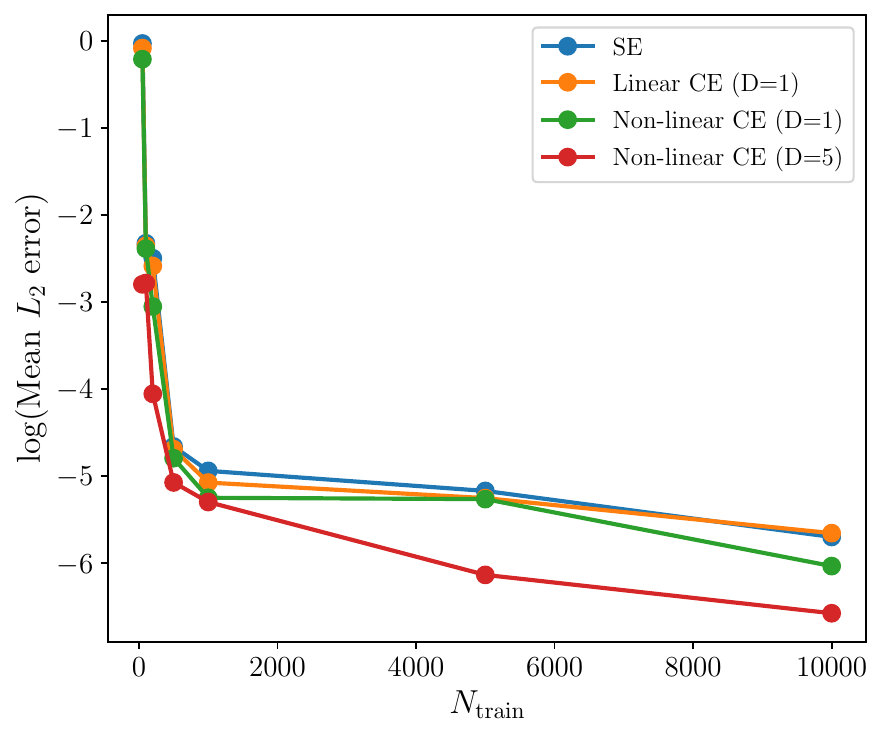}
\par\end{centering}
\caption{Variation of the test $L_2$ errors with varying size of the input dataset ($N_\text{train}$.)}\label{fig:1d_oneint_abaltion}
\end{figure}

Figure~\ref{fig:1d_oneint_abaltion} shows how the relative $L_2$ error distribution changes with different numbers of training samples ($N_\text{train}$). The errors drop sharply as $N_\text{train}$ increases from 50 to 800, after which the improvement becomes minimal once $N_\text{train}$ exceeds 5000. Therefore, for all remaining examples, we use $N_\text{train} = 5000$.

\begin{figure}[!hbt]
    \centering
    \begin{subfigure}[b]{0.23\textwidth}
        \centering
        \includegraphics[width=\textwidth]{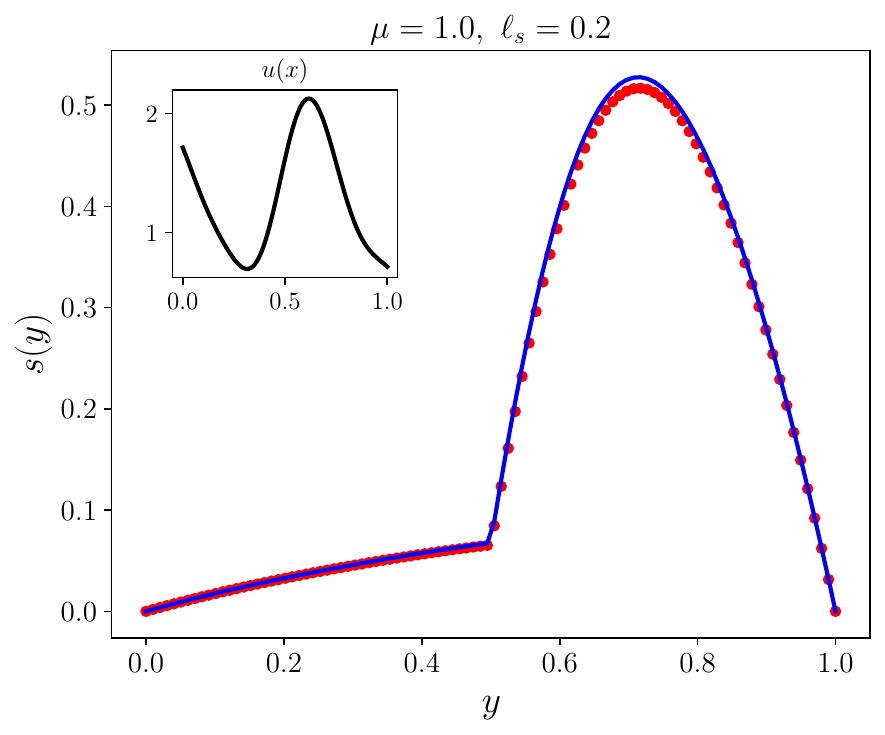}
        \caption{}
    \end{subfigure}
    \hspace{2.0pt}
    \begin{subfigure}[b]{0.23\textwidth}
        \centering
        \includegraphics[width=\textwidth]{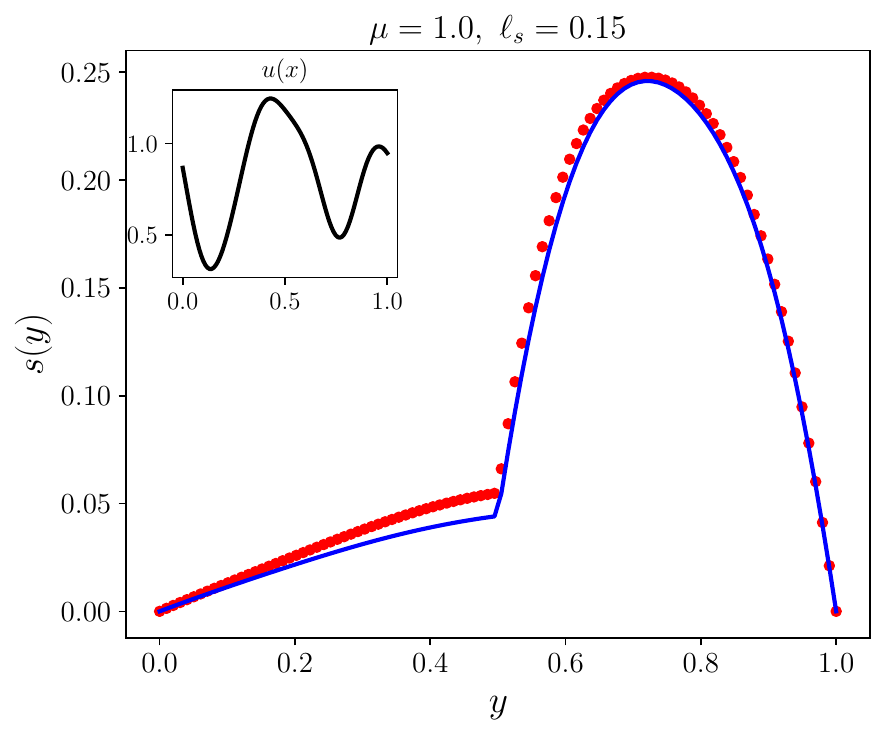}
        \caption{}
    \end{subfigure} 
    \hspace{2.0pt}
    \begin{subfigure}[b]{0.23\textwidth}
        \centering
        \includegraphics[width=\textwidth]{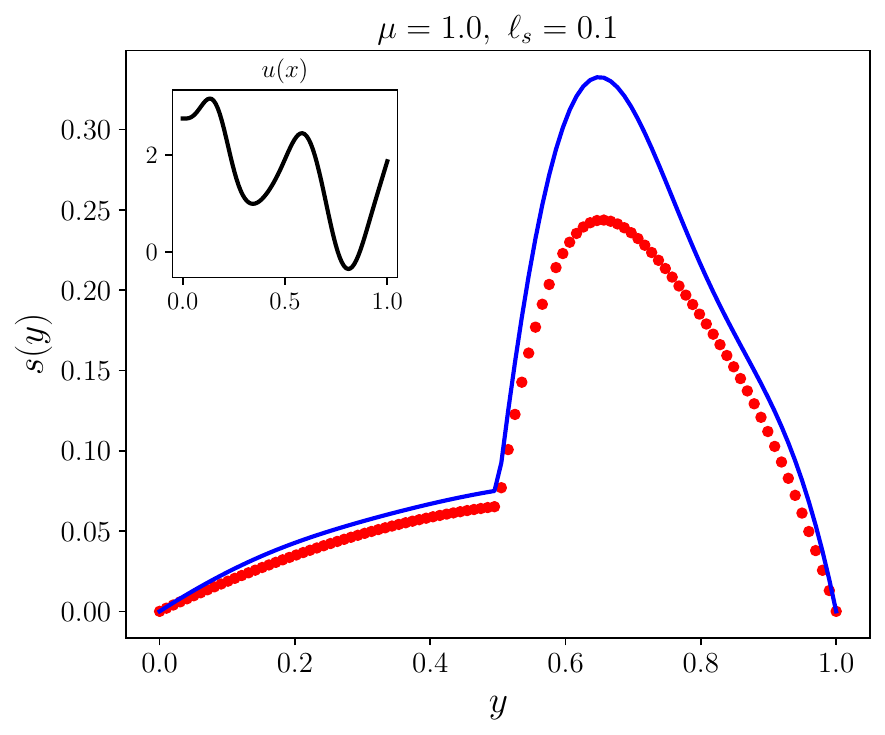}
        \caption{}
    \end{subfigure} 
    \hspace{2.0pt}
    \begin{subfigure}[b]{0.23\textwidth}
        \centering
        \includegraphics[width=\textwidth]{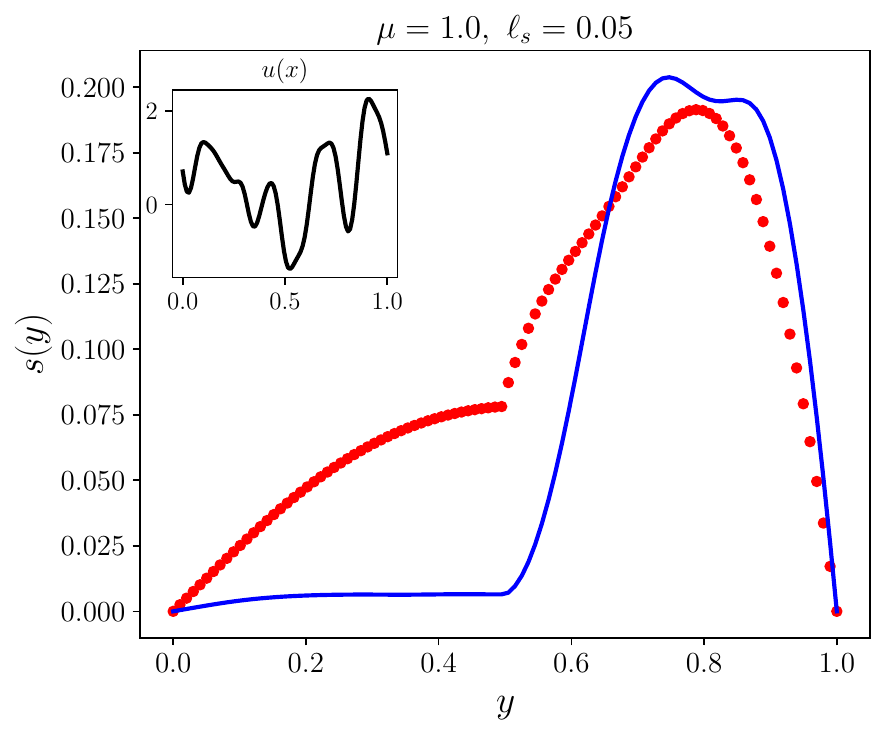}
        \caption{}
    \end{subfigure} \\

    \centering
    \begin{subfigure}[b]{0.23\textwidth}
        \centering
        \includegraphics[width=\textwidth]{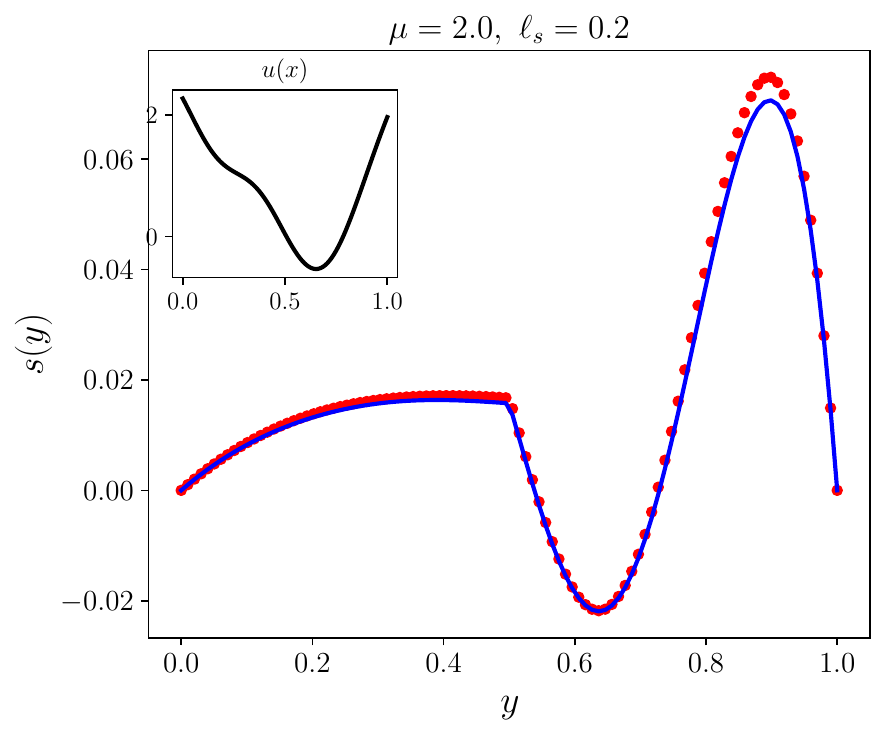}
        \caption{}
    \end{subfigure}
    \hspace{2.0pt}
    \begin{subfigure}[b]{0.23\textwidth}
        \centering
        \includegraphics[width=\textwidth]{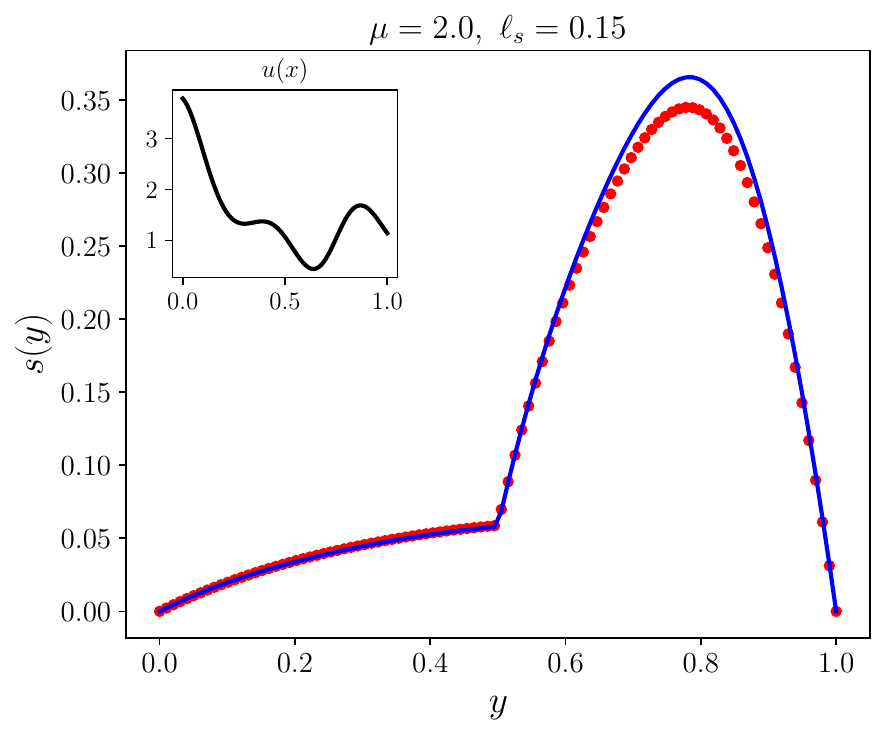}
        \caption{}
    \end{subfigure} 
    \hspace{2.0pt}
    \begin{subfigure}[b]{0.23\textwidth}
        \centering
        \includegraphics[width=\textwidth]{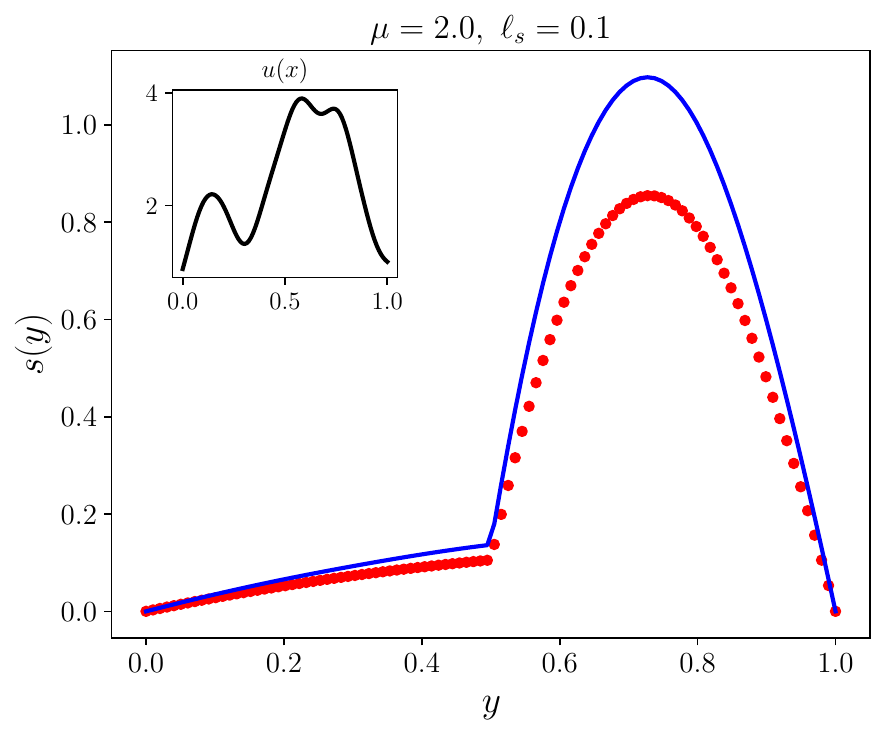}
        \caption{}
    \end{subfigure} 
    \hspace{2.0pt}
    \begin{subfigure}[b]{0.23\textwidth}
        \centering
        \includegraphics[width=\textwidth]{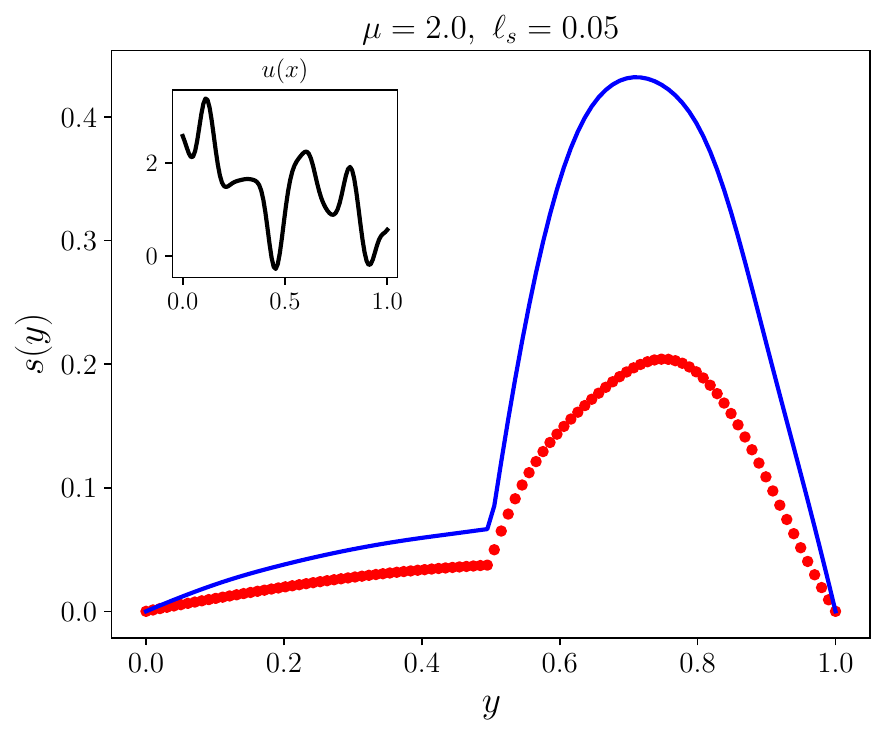}
        \caption{}
    \end{subfigure} \\

    \centering
    \begin{subfigure}[b]{0.23\textwidth}
        \centering
        \includegraphics[width=\textwidth]{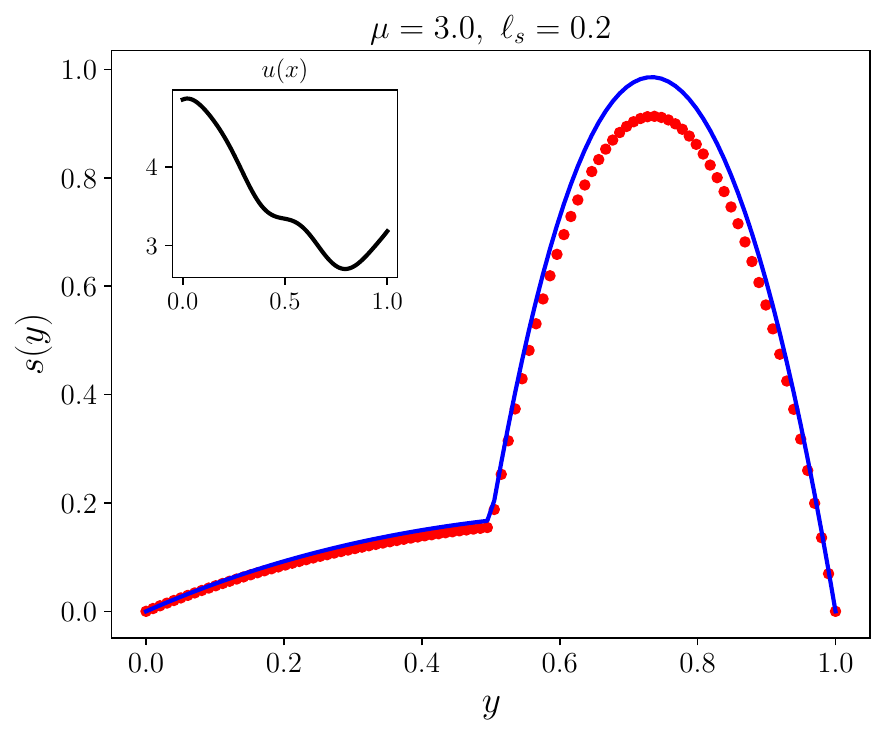}
        \caption{}
    \end{subfigure}
    \hspace{2.0pt}
    \begin{subfigure}[b]{0.23\textwidth}
        \centering
        \includegraphics[width=\textwidth]{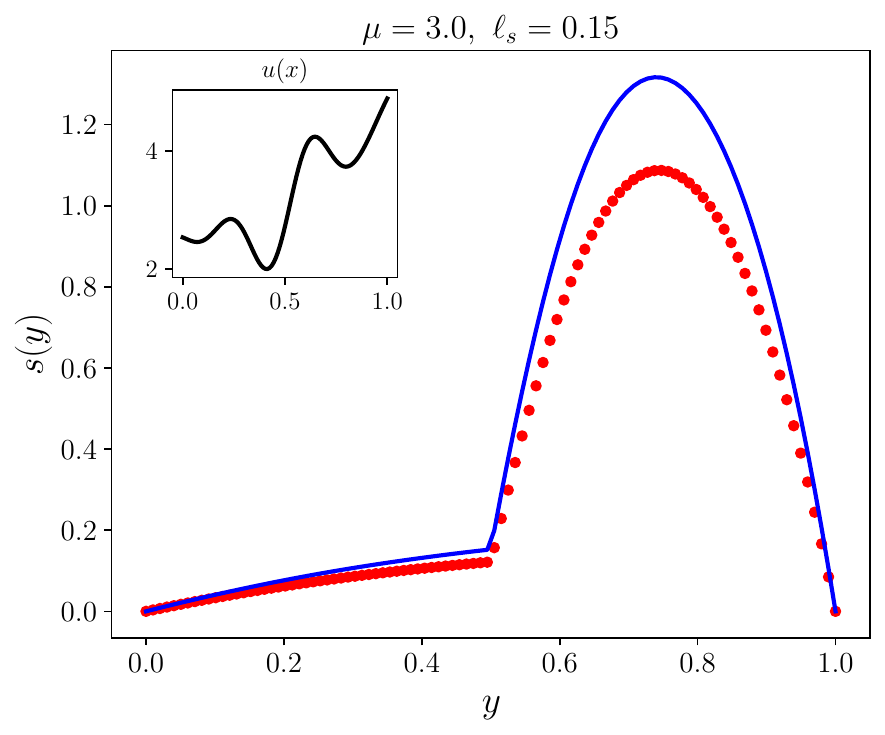}
        \caption{}
    \end{subfigure} 
    \hspace{2.0pt}
    \begin{subfigure}[b]{0.23\textwidth}
        \centering
        \includegraphics[width=\textwidth]{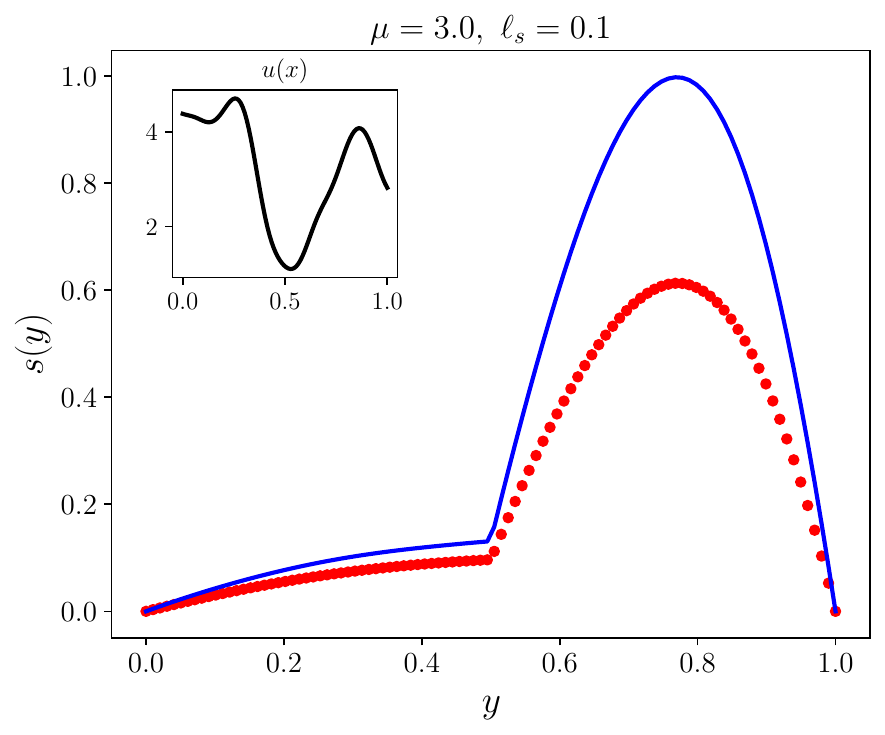}
        \caption{}
    \end{subfigure} 
    \hspace{2.0pt}
    \begin{subfigure}[b]{0.23\textwidth}
        \centering
        \includegraphics[width=\textwidth]{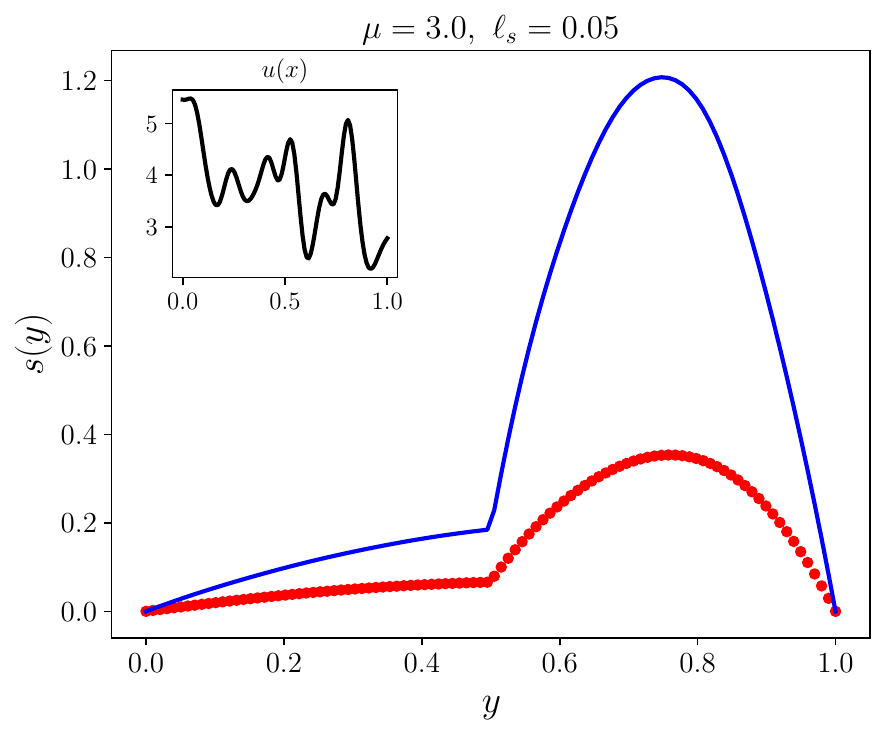}
        \caption{}
    \end{subfigure} \\

    \caption{Prediction of the $\phi$-DeepONet model on out-of-distribution test samples for the 1D problem with one interface (Section~\ref{sec:1d_oneint}). The model is trained on input functions sampled from a $\mathcal{GP}$ with mean $\mu = 1.0$ and length scale $l_s = 0.2$ (Eqs.~\eqref{eq:gp_inputs_a} and \eqref{eq:gp_inputs_b}), and tested on several other combinations of $\mu$ and $l_s$. Across the rows, $\mu$ is varied, and across the columns, $l_s$ is varied. The corresponding input functions are also shown in each of the subplots.}
    \label{fig:1d_oneint_ood_predictions}
\end{figure}

\textbf{Out-of-distribution (OOD) prediction capability:} We also study the performance of the $\phi$-DeepONet framework on various out-of-distribution test samples. 
This is done by varying the mean $\mu$ and length scale $l_s$ of the $\mathcal{GP}$ from which the input functions are sampled. Figure~\ref{fig:1d_oneint_ood_predictions} shows predictions for increasing values of $\mu$ (across rows) and decreasing values of $l_s$ (across columns). The model was trained using input functions with $\mu = 1.0$ and $l_s = 0.2$. Note that we do not include tests with $l_s > 0.2$ because larger length scales make the solution smoother and make the problem artificially easier. We observe that for a fixed $l_s$, the predictions remain fairly stable across different values of $\mu$. However, the performance degrades as $l_s$ decreases. The model performs reasonably well for $0.2 \geq l_s \geq 0.15$, but its accuracy drops rapidly beyond this threshold. Interestingly, the model's performance is almost unaffected by changes in $\mu$, regardless of the choice of $l_s$. Previous studies have shown that incorporating Fourier feature encodings in the input layer can improve the OOD generalization of neural operators~\cite{wang2021learning}. In this work, however, we intentionally avoid such modifications in order to assess the intrinsic OOD capability of the proposed framework. For more challenging OOD settings, we recommend augmenting the model with such techniques.

\subsection{One-dimensional problem with multiple interfaces}\label{sec:1d_fourint}

Next, we increase the complexity of the previous problem by adding more interfaces. Instead of a single interface, we now consider four interfaces. The problem is redefined as follows. The domain is $\Omega = [0,1]$ with interfaces located at $\Gamma_{\text{int}_1}=0.2$, $\Gamma_{\text{int}_2}=0.4$, $\Gamma_{\text{int}_3}=0.6$, and $\Gamma_{\text{int}_4}=0.8$, which divide the domain into five subdomains:
\[
\Omega_1=[0,\Gamma_{\text{int}_1}], \quad
\Omega_2=[\Gamma_{\text{int}_1},\Gamma_{\text{int}_2}], \quad
\Omega_3=[\Gamma_{\text{int}_2},\Gamma_{\text{int}_3}], \quad
\Omega_4=[\Gamma_{\text{int}_3},\Gamma_{\text{int}_4}], \quad
\Omega_5=[\Gamma_{\text{int}_4},1],
\]
with $\Omega = \Omega_1 \cup \Omega_2 \cup \Omega_3 \cup \Omega_4 \cup \Omega_5$.  The goal is to approximate the operator $G: u(x) \rightarrow s(y)$, where $u(x)$ is again sampled from a GRF with mean $\mu = 1.0$ and length scale $l_s = 0.2$. The learned operator should approximate the solution of the PDE:

\begin{align}
\label{eq:1d_fourint}
  \begin{split}
      \frac{d}{dy}\left(\kappa_q \frac{d s_q}{dy}\right) &= u_q \quad \text{in } \Omega_q, \\
      s_1 &= 0 \quad \text{at } y=0, \\
      s_5 &= 0 \quad \text{at } y=1, \\
      \llbracket s \rrbracket &= 0 \quad \text{at } y=\Gamma_{\text{int}_i}, \quad i = 1,2,3,4, \\
      \left\llbracket \kappa \frac{d s}{dy} \right\rrbracket &= 0 \quad \text{at } y=\Gamma_{\text{int}_i}, \quad i = 1,2,3,4.
  \end{split}
\end{align}
The material constants for the five subdomains are set as $\kappa_1 = 2$, $\kappa_2 = 0.1$, $\kappa_3 = 0.5$, $\kappa_4 = 2$, and $\kappa_5 = 0.7$. While the output function $s(y)$ is continuous, its derivative is not due to the piecewise constant material parameters. We use $N_\text{train} = 5000$ and $N_\text{test} = 500$.

Table~\ref{tab:1d_fourint_errors} shows the relative $L_2$ errors on the test dataset for the different versions of the $\phi$-DeepONet framework, along with results from the physics-informed DeepONet and IONet frameworks. As before, the standard DeepONet performs worse than all $\phi$-DeepONet and IONet variants. Comparing the $\phi$-DeepONet models, we observe that the simple SE model performs better than the CE model. This is interesting, and while the exact reason is not clear, we believe it may be related to the fact that explicitly specifying the latent dimension leads to a more effective domain decomposition embedding in this case. However, this may not always hold true, as seen in later examples. For the non-linear models, we see that the errors improve from $\mathcal{O}(10^{-1})$ to $\mathcal{O}(10^{-2})$ when the latent dimension increases from $D=1$ to $D=2$. Beyond $D=3$, the difference in accuracy is small, while the training cost continues to increase, which becomes important for large-scale problems. Therefore, for the results presented in this section, we use the non-linear SE model with latent dimension $D=3$. Regarding the computational costs, while IONet leads to marginally better approximation than $\phi-$DeepONet variants, the cost of training scales up to almost 2.7 times that of the $\phi-$DeepONet.

\begin{table}[h!]
\centering
\begin{tabular}{l c c c c c c c c c}
\toprule
& \textbf{SE} 
& \textbf{CE} 
& \multicolumn{5}{c}{\textbf{Non-linear CE}} 
& \textbf{DeepONet}
& \textbf{IONet}\\
\cmidrule(lr){4-8}
&  &  & $D=1$ & $D=2$ & $D=3$ & $D=4$ & $D=5$ &  \\
\midrule
\textbf{Rel.~$L_2$ error}
& $7.41\text{e-}2$
& $3.14\text{e-}1$
& $2.13\text{e-}1$
& $7.85\text{e-}2$
& $3.43\text{e-}2$
& $3.01\text{e-}2$
& $1.18\text{e-}2$
& $2.29\text{e}0$
& $9.81\text{e-}3$ \\
\textbf{Cost}
& 0.98
& 1.0
& 1.0
& 0.99
& 1.0
& 1.12
& 1.14
& 1.39
& 2.71 \\
\bottomrule
\end{tabular}
\caption{Relative $L_2$ errors on the test set for the various $\phi$-DeepONet frameworks (along with the standard physics-informed DeepONet and IONet models) for the 1D problem with four interfaces (Section~\ref{sec:1d_fourint}).}\label{tab:1d_fourint_errors}
\end{table}

\begin{figure}[!hbt]
    \centering
    \begin{subfigure}[b]{0.7\textwidth}
        \centering
        \includegraphics[width=\textwidth]{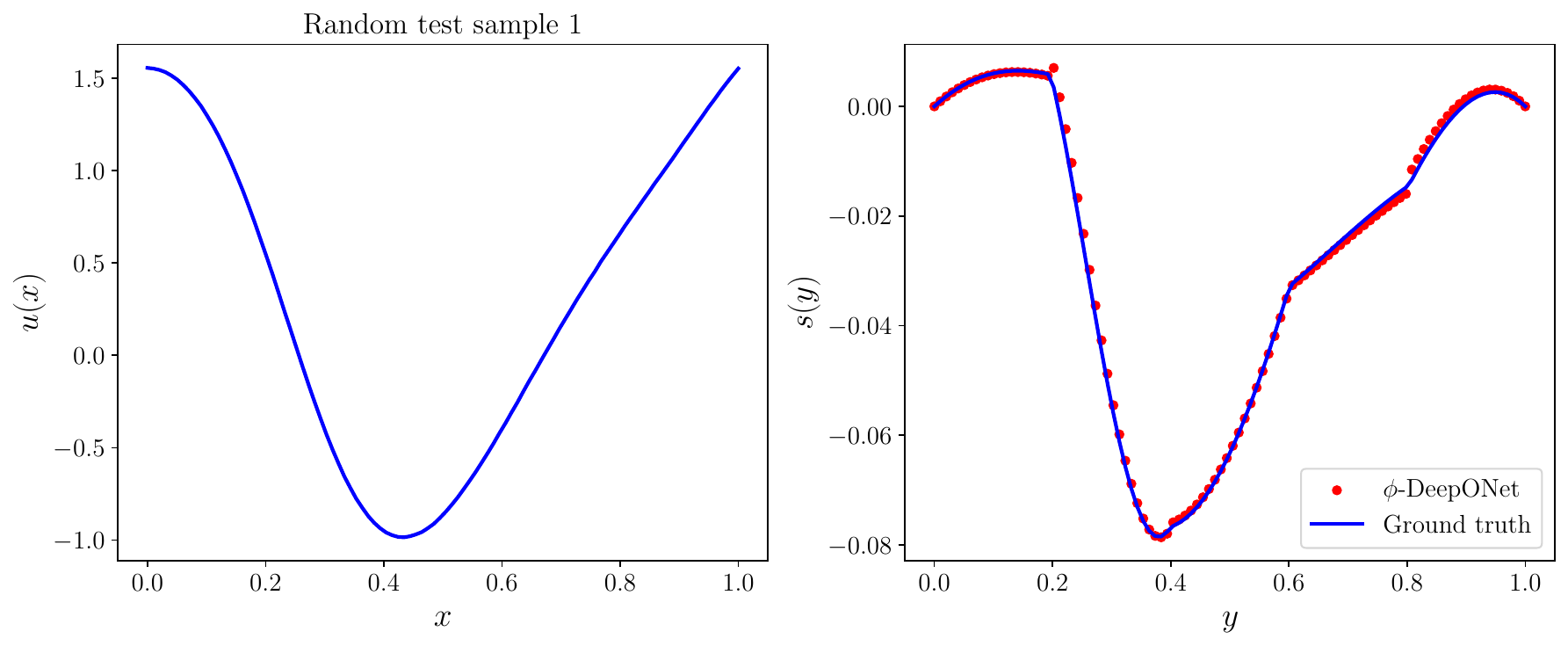}
        \caption{}
        \label{fig:1d_fourint_compare_test1}
    \end{subfigure}\\
    \centering
    \begin{subfigure}[b]{0.7\textwidth}
        \centering
        \includegraphics[width=\textwidth]{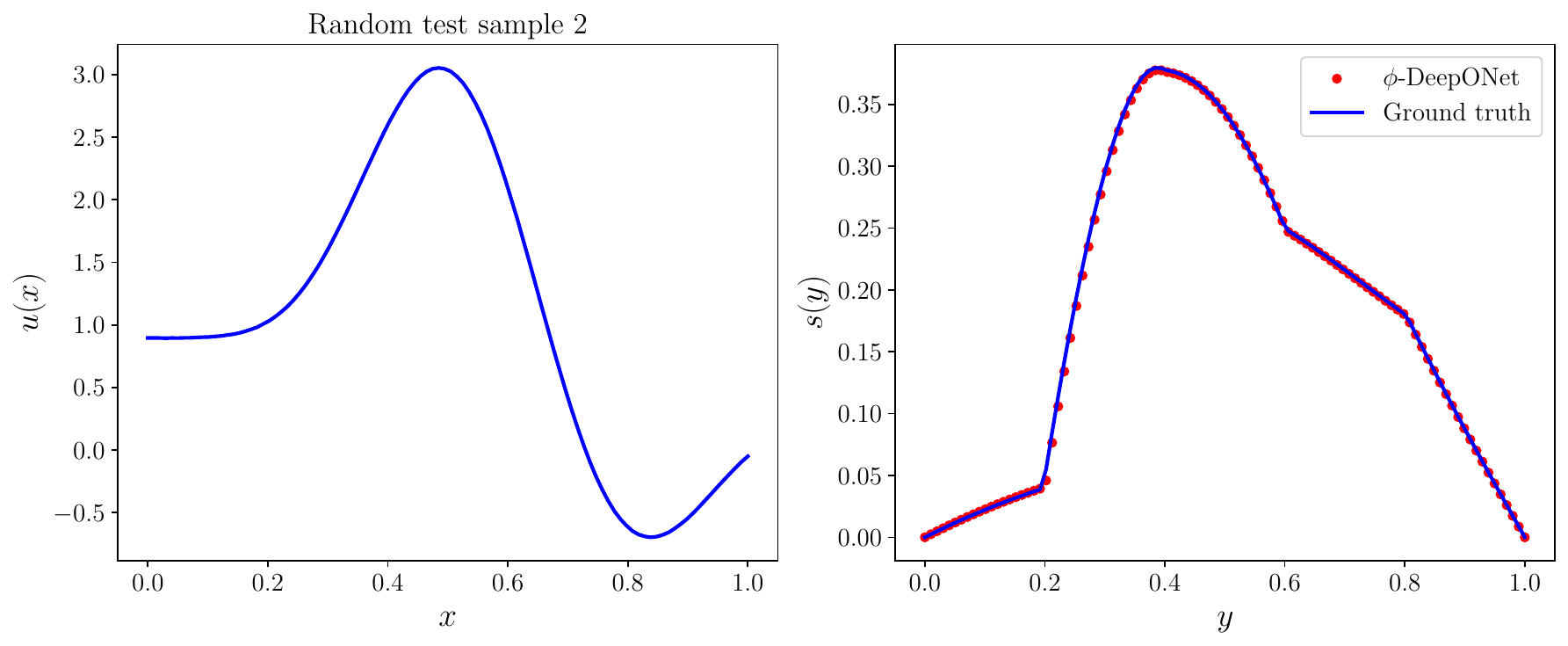}
        \caption{}
        \label{fig:1d_fourint_compare_test2}
    \end{subfigure}\\
    \centering
    \begin{subfigure}[b]{0.4\textwidth}
        \centering
        \includegraphics[width=\textwidth]{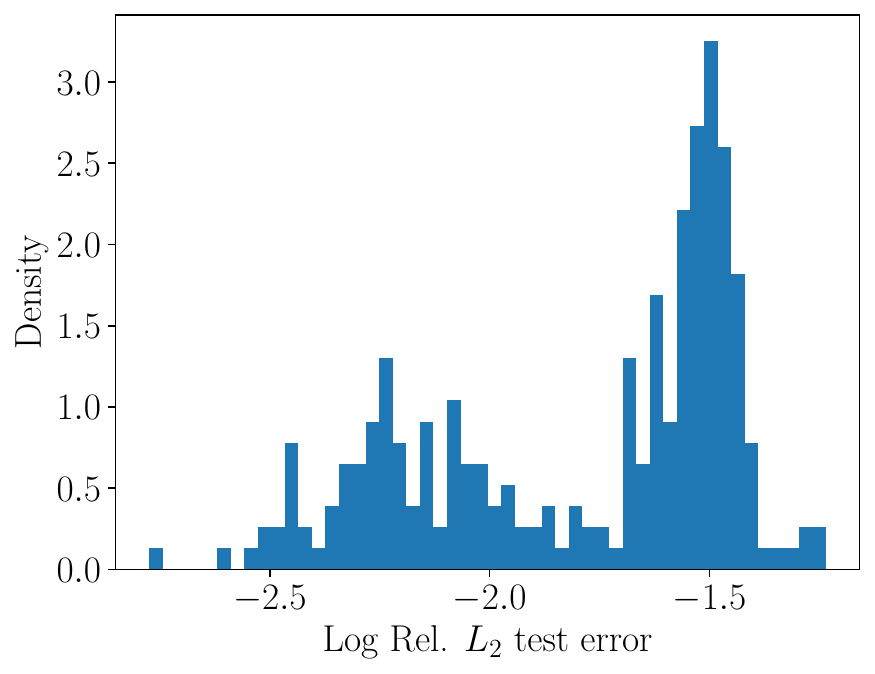}
        \caption{}
        \label{fig:1d_fourint_errorplot}
    \end{subfigure}\\
    \caption{1D multiple-interface problem: performance of the $\phi$-DeepONet framework on two random test samples (subfigures (a) and (b)), along with the distribution of the test errors (subfigure (c)).}
    \label{fig:1d_fourint_examples_with_convergence}
\end{figure}

Figure~\ref{fig:1d_fourint_examples_with_convergence} shows the predictions of the trained $\phi$-DeepONet framework (non-linear CE with $D=3$) on two randomly selected test samples. The distribution of the test errors is shown in Figure~\ref{fig:1d_fourint_errorplot}. We observe that the predicted solutions match the reference (ground-truth) solutions closely, similar to the previous case, with the reference obtained from the same numerical solver. Figure~\ref{fig:1d_fourint_errorplot} shows the distribution of the mean relative $L_2$ test errors. Interestingly, the error distribution exhibits a bimodal structure. While the precise cause of this behavior is not yet fully understood, a plausible explanation is related to the role of the categorical embedding. In problems with multiple interfaces, the CE component is required to learn a more complex latent partition of the solution space in addition to the operator mapping itself. This added representational burden may give rise to two distinct regimes of predictive performance: one in which the learned embedding aligns well with the underlying interface structure, and another in which the representation is less effective. Notably, such bimodal behavior is not observed for the SE-based models, suggesting that this phenomenon is unlikely to arise solely from the data distribution. Instead, it appears to stem from the interaction between the CE parameterization and the problem-dependent interface complexity.

\begin{figure}[!hbt]
    \centering
    \begin{subfigure}[b]{0.3\textwidth}
        \centering
        \includegraphics[width=\textwidth]{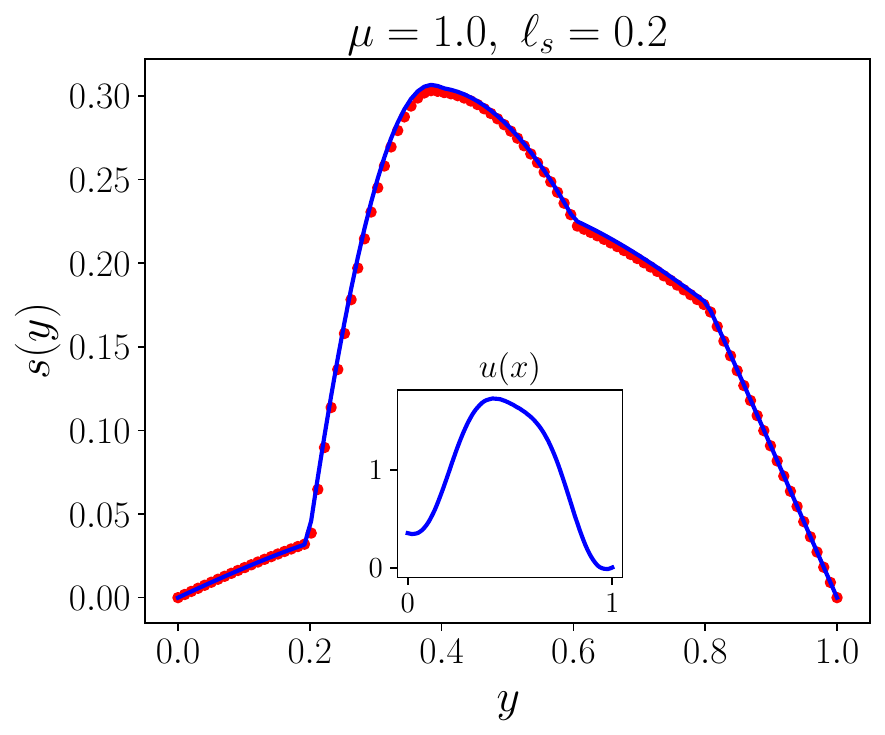}
        \caption{}
        \label{fig:1d_fourint_ood1a}
    \end{subfigure}
    \hspace{2.0pt}
    \begin{subfigure}[b]{0.3\textwidth}
        \centering
        \includegraphics[width=\textwidth]{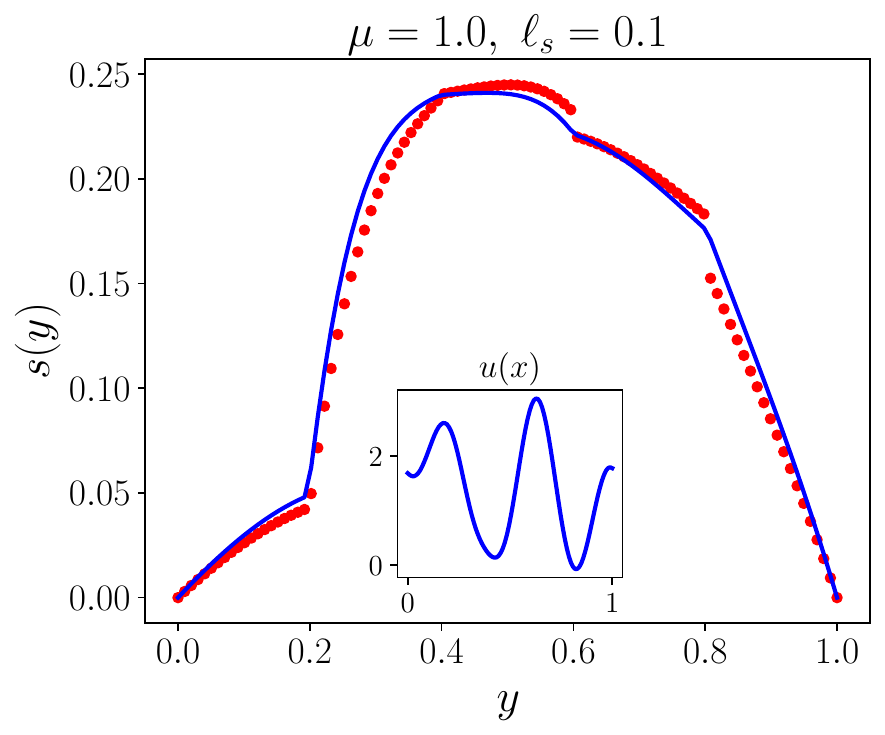}
        \caption{}
        \label{fig:1d_fourint_ood1b}
    \end{subfigure} 
    \hspace{2.0pt}
    \begin{subfigure}[b]{0.3\textwidth}
        \centering
        \includegraphics[width=\textwidth]{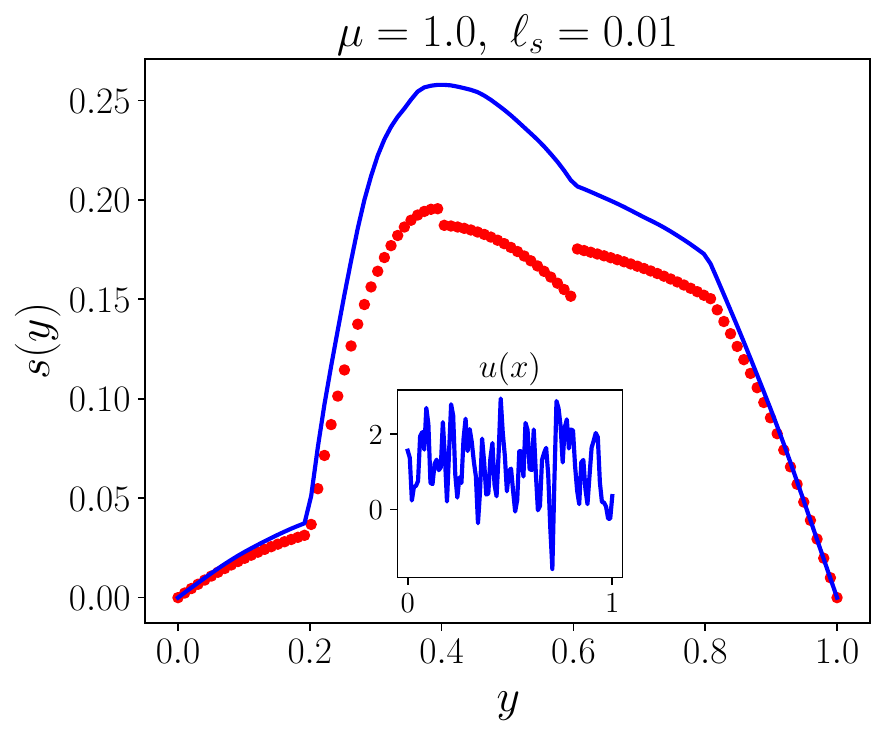}
        \caption{}
        \label{fig:1d_fourint_ood1c}
    \end{subfigure} \\

    \centering
    \begin{subfigure}[b]{0.3\textwidth}
        \centering
        \includegraphics[width=\textwidth]{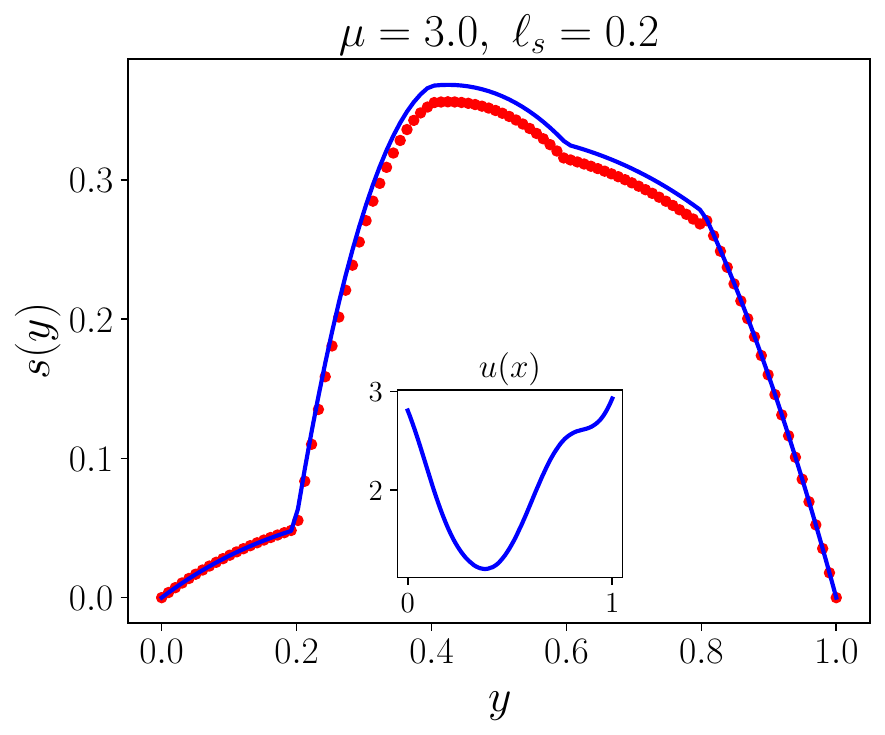}
        \caption{}
        \label{fig:1d_fourint_ood2a}
    \end{subfigure}
    \hspace{2.0pt}
    \begin{subfigure}[b]{0.3\textwidth}
        \centering
        \includegraphics[width=\textwidth]{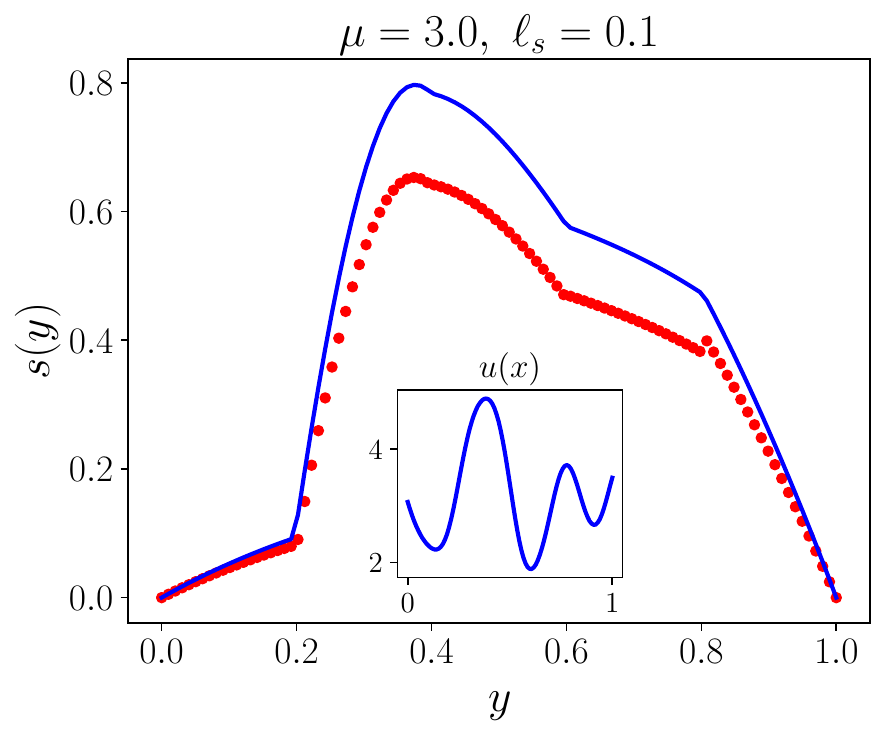}
        \caption{}
        \label{fig:1d_fourint_ood2b}
    \end{subfigure} 
    \hspace{2.0pt}
    \begin{subfigure}[b]{0.3\textwidth}
        \centering
        \includegraphics[width=\textwidth]{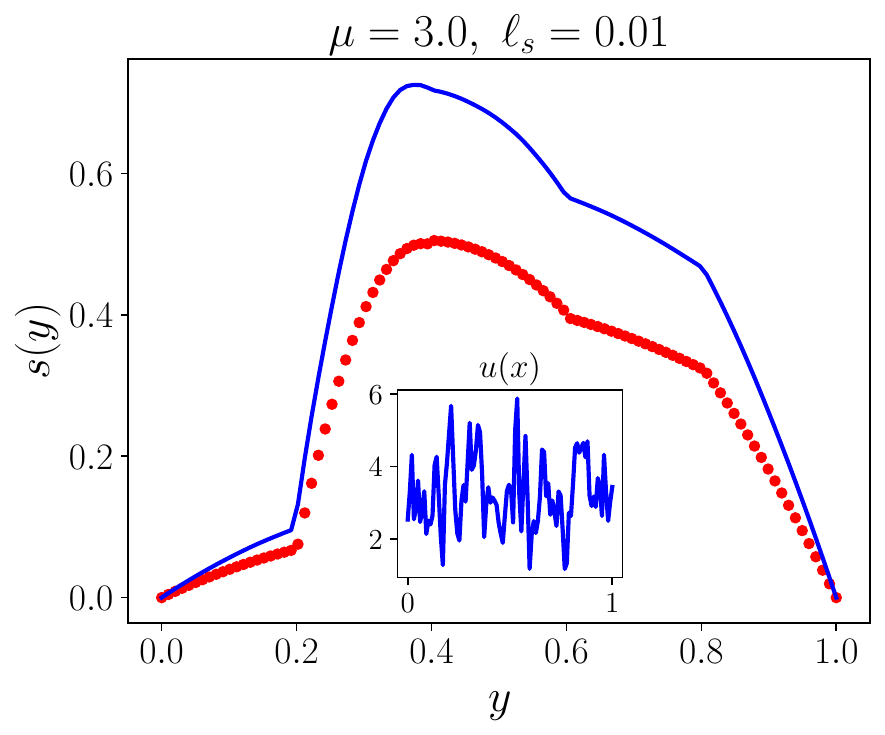}
        \caption{}
        \label{fig:1d_fourint_ood2c}
    \end{subfigure} \\

    \centering
    \begin{subfigure}[b]{0.3\textwidth}
        \centering
        \includegraphics[width=\textwidth]{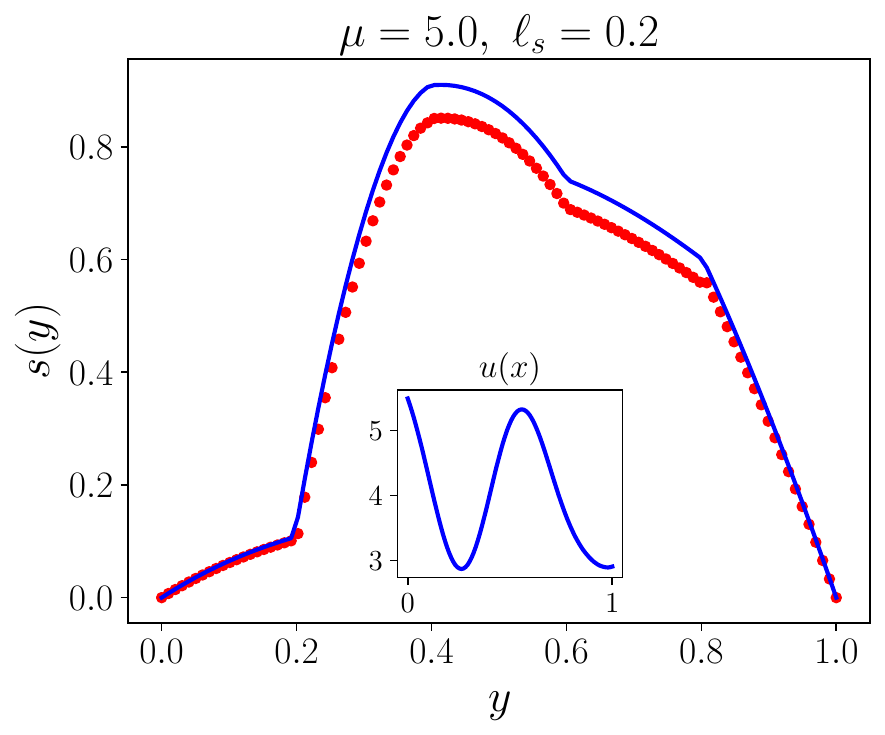}
        \caption{}
        \label{fig:1d_fourint_ood3a}
    \end{subfigure}
    \hspace{2.0pt}
    \begin{subfigure}[b]{0.3\textwidth}
        \centering
        \includegraphics[width=\textwidth]{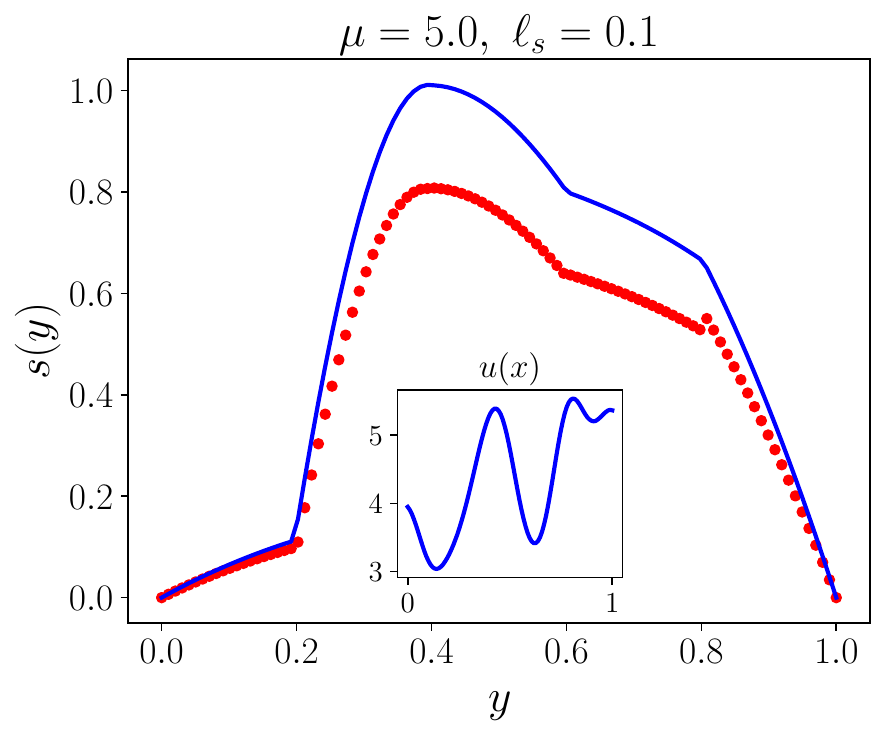}
        \caption{}
        \label{fig:1d_fourint_ood3b}
    \end{subfigure} 
    \hspace{2.0pt}
    \begin{subfigure}[b]{0.3\textwidth}
        \centering
        \includegraphics[width=\textwidth]{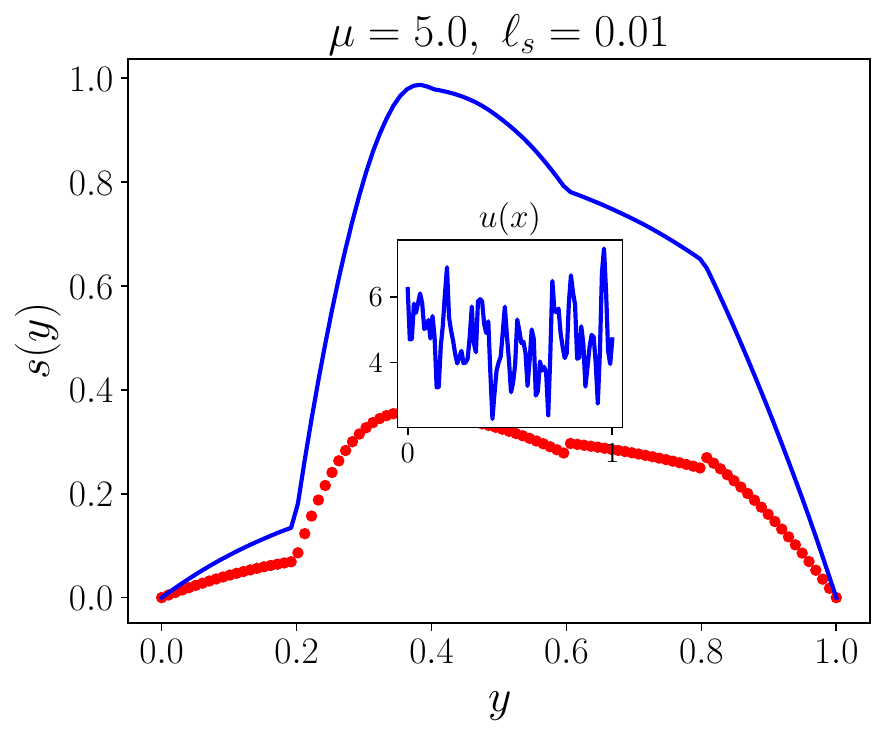}
        \caption{}
        \label{fig:1d_fourint_ood3c}
    \end{subfigure} \\

    \caption{Prediction of the $\phi$-DeepONet model on out-of-distribution test samples for the 1D problem with four interfaces (Section~\ref{sec:1d_fourint}). The model is trained on input functions sampled from a $\mathcal{GP}$ with mean $\mu = 1.0$ and length scale $l_s = 0.2$ (Eqs.~\eqref{eq:gp_inputs_a} and \eqref{eq:gp_inputs_b}), and tested on several other combinations of $\mu$ and $l_s$. Across the rows, $\mu$ is varied, and across the columns, $l_s$ is varied. The corresponding input functions fed to the model is also shown in each of the subplots.}
    \label{fig:1d_fourint_ood_predicitons}
\end{figure}

\textbf{Out-of-distribution prediction (OOD) capability:} We again analyze the out-of-distribution prediction capability of the $\phi$-DeepONet framework. The input functions in the training set (sampled from a $\mathcal{GP}$ with $\mu = 1.0$ and $l_s = 0.2$) are different from the input functions used in the test set, where both $\mu$ and $l_s$ are varied. Figure~\ref{fig:1d_fourint_ood_predicitons} shows the performance of the trained $\phi$-DeepONet model on these out-of-distribution samples. We observe a similar trend as before: the model's predictions remain stable across different values of the mean $\mu$, but the accuracy worsens as the length scale $l_s$ decreases. As earlier, we do not test with larger $l_s$ than those used during training. As previously stated, incorporating modifications such as Fourier feature encodings in the input layer~\cite{wang2021learning} can further enhance the OOD generalization capability of neural operators.

\subsection{Two-dimensional problem}\label{sec:2d}

Next, we consider a two-dimensional Poisson equation on a discontinuous domain with a forcing function. This type of problem setup is common in many areas of science and engineering, for example heat conduction in layered materials, groundwater flow in layered soils, and diffusion in multi-phase materials to name a few. We consider a square domain $\Omega = [0,1]\times[0,1]$ with an interface $\Gamma_\text{int} = \{\boldsymbol{x} : x_2 = x_1\}$. This interface divides the domain into two non-overlapping subdomains, $\Omega_1 = \{\boldsymbol{x} : x_2 \le x_1\}$ and $\Omega_2 = \{\boldsymbol{x} : x_2 > x_1\}$, such that $\Omega = \Omega_1 \cup \Omega_2$. We aim to approximate the non-linear operator $\mathcal{G} : u(\boldsymbol{x}) \mapsto s(\boldsymbol{y})$ that satisfies the PDE in each region $q=1,2$:
\begin{align}
  \begin{split}
    \nabla\cdot(\kappa_q\nabla s_q) &= u_q \qquad\text{in } \Omega_q,\\
    s_q &= 0 \qquad\text{on } \partial\Omega^\text{d}_q,\\
    \llbracket s \rrbracket &= 0 \qquad\text{on } \Gamma_\text{int},\\
    \llbracket \kappa \nabla s \rrbracket \cdot \mathbf{n}_2 &= 0 \qquad\text{on } \Gamma_\text{int},
  \end{split}
  \label{eq:2d}
\end{align}
where zero Dirichlet boundary conditions are prescribed on parts of the external boundary:
\[
\partial\Omega^\text{d}_1 = \{\boldsymbol{x}: x = 1 \cup y = 0\}, \qquad
\partial\Omega^\text{d}_2 = \{\boldsymbol{x}: x = 0 \cup y = 1\}.
\]
The input function $u(\boldsymbol{x})$ is sampled from a $\mathcal{GP}$ with $\mu = 1.0$ and $l_s = 0.2$. The material parameters are set to $\kappa_1 = 1.0$ and $\kappa_2 = 0.2$. We train several variants of the $\phi$-DeepONet model and compare their performance with numerical solutions obtained from a finite difference method.

\begin{figure}[!hbt]
    \centering
    \begin{subfigure}[b]{0.9\textwidth}
        \centering
        \includegraphics[width=\textwidth]{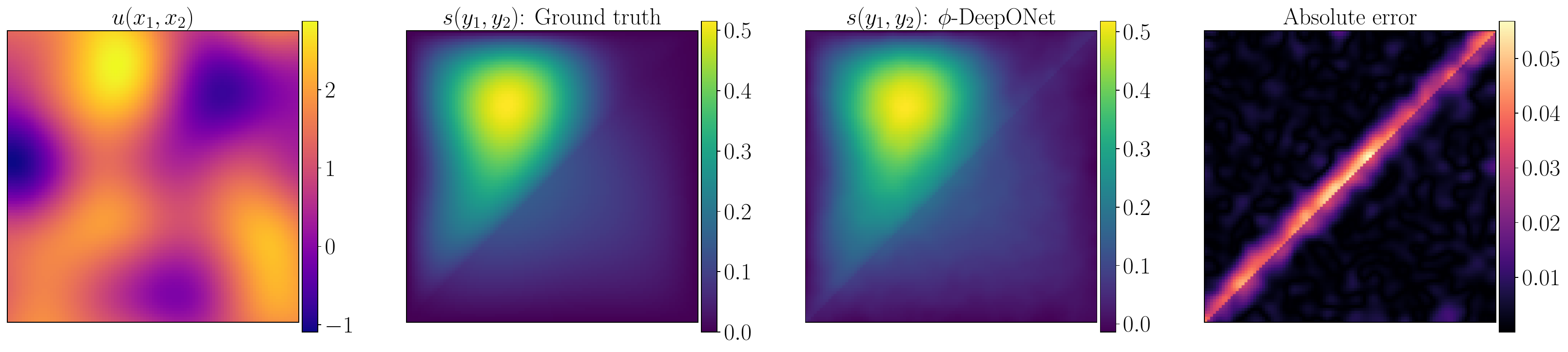}
        \caption{}
        \label{fig:2d_oneint_testexample_1}
    \end{subfigure}\\
    \centering
    \begin{subfigure}[b]{0.9\textwidth}
        \centering
        \includegraphics[width=\textwidth]{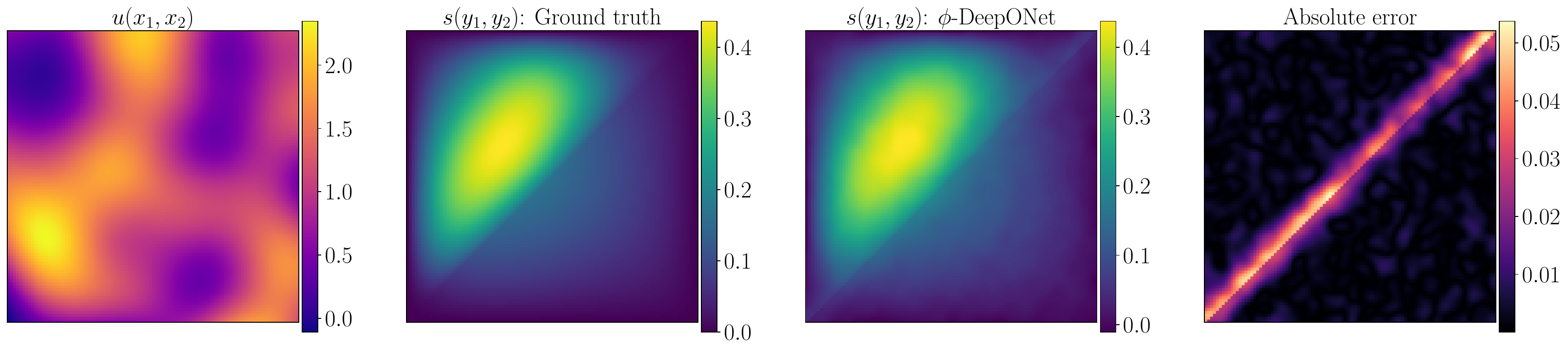}
        \caption{}
        \label{fig:2d_oneint_testexample_2}
    \end{subfigure}\\
    \centering
    \begin{subfigure}[b]{0.4\textwidth}
        \centering
        \includegraphics[width=\textwidth]{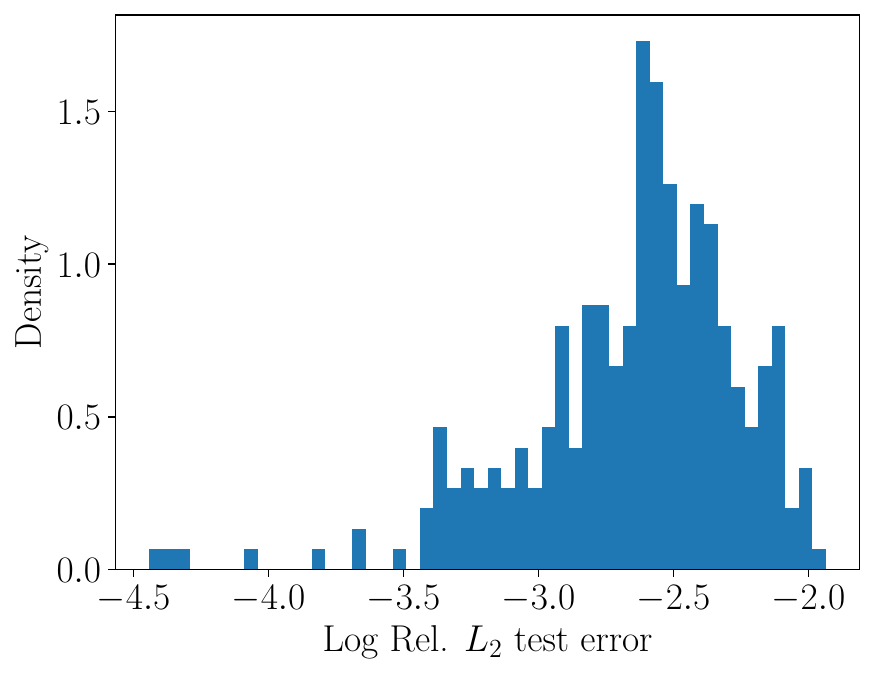}
        \caption{}
        \label{fig:2d_oneint_discontu_errorplot}
    \end{subfigure}\\
    \caption{Performance of the $\phi$-DeepONet (non-linear CE, $D = 3$) framework on two random test samples for the 2D problem (subfigures (a) and (b)), along with the distribution of the test errors (sub-figure (c)).}
    \label{fig:2d_oneint}
\end{figure}

Table~\ref{tab:2d} shows the mean relative $L_2$ errors for the various $\phi$-DeepONet frameworks on the test set, alongside the physics-informed DeepONet and IONet frameworks. Unlike the previous case (Table~\ref{tab:1d_fourint_errors}), the CE models (for $D > 1$) achieve better accuracy than the SE models. However, similar to before, for $D \geq 3$ the performance across models improves only marginally. The standard DeepONet produces errors at least two orders of magnitude higher than the non-linear CE $\phi$-DeepONet models. For similar accuracy, the IONet framework needs approximately 2.5 times the computational cost of the $\phi-$DeepOet models. Figure~\ref{fig:2d_oneint} shows the predictions of the trained $\phi$-DeepONet model (non-linear CE with $D=3$) for two random test samples, alongside the corresponding ground-truth solutions and absolute error plots. The predictions align well with the ground truth, with most of the error concentrated around the interface. Figure~\ref{fig:2d_oneint_discontu_errorplot} presents the distribution of the mean relative $L_2$ errors over the full test set.

\begin{table}[h!]
\centering
\begin{tabular}{l c c c c c c c c c}
\toprule
& \textbf{SE} 
& \textbf{CE} 
& \multicolumn{5}{c}{\textbf{Non-linear CE}} 
& \textbf{DeepONet}
& \textbf{IONet} \\
\cmidrule(lr){4-8}
&  &  & $D=1$ & $D=2$ & $D=3$ & $D=4$ & $D=5$ &  \\
\midrule
\textbf{Rel.~$L_2$ error}
& $5.38\text{e-}2$
& $3.21\text{e-}2$
& $4.68\text{e-}2$
& $9.37\text{e-}3$
& $4.46\text{e-}3$
& $3.22\text{e-}3$
& $2.71\text{e-}3$
& $5.13\text{e-}1$
& $1.05\text{e-}3$ \\
\textbf{Cost}
& 0.91
& 1.02
& 1.0
& 1.14
& 1.10
& 1.27
& 1.33
& 1.81
& 2.55 \\
\bottomrule
\end{tabular}
\caption{Relative $L_2$ errors on the test set for the various $\phi$-DeepONet frameworks (along with the standard physics-informed DeepONet and IONet models) for the 2D problem (Section~\ref{sec:2d}).}\label{tab:2d}
\end{table}

\section{Numerical Examples with Discontinuous Inputs}\label{sec:numerical_examples_discontint}

We now test the performance of the $\phi-$DeepONet architectures in problems where, in addition to discontinuous material parameters $\kappa$, the input functions $u(x)$ are also piecewise continuous. This setting requires the use of multiple branch networks, with one branch network $br_q$ assigned to each subdomain $\Omega_q$. For all problems considered in this section, each input function $u(x)$ is constructed as a collection of subdomain-specific functions $u_q(x)$. These functions are generated from the same Gaussian process across all subdomains, as defined in Eq.~\eqref{eq:gp_inputs_a}.

\subsection{One-dimensional problem}\label{sec:1d_fourint_discoituous}

We begin with the 1D problem discussed earlier in Section~\ref{sec:1d_fourint}, but now consider the case where the input function $u(x)$ is also piecewise discontinuous. Specifically, for each subdomain $\Omega_q$, the input function $u_q(x)$ is piecewise continuous within that subdomain. The material constants in the five subdomains are set as $\kappa_1 = 1.0$, $\kappa_2 = 5.0$, $\kappa_3 = 1.0$, $\kappa_4 = 2.0$, and $\kappa_5 = 5.0$. The objective is to approximate the mapping from the composite input function $\tilde{u}(x)$ to the composite output function $\tilde{s}(y)$, given by the operator $\mathcal{G} : \tilde{u}(x) \mapsto \tilde{s}(y)$. For this problem, we use $N_{\text{train}} = 10{,}000$ training samples and $N_{\text{test}} = 1{,}000$ test samples.

Table~\ref{tab:1d_fourint_discont_errors} reports the relative $L_2$ errors on the test dataset for the three variants of the $\phi$-DeepONet framework. As expected, due to the increased complexity arising from discontinuities in both the input and output functions, as well as in the material parameters, the basic SE and CE versions of $\phi$-DeepONet achieve relative $L_2$ errors on the order of $\mathcal{O}(10^{-1})$. This level of accuracy remains similar for the non-linear variants when the latent dimension is small. Once the latent dimension reaches $D \geq 3$, the relative $L_2$ errors improve to the order of $\mathcal{O}(10^{-2})$. And similar to previous trends, while maintaining similar levels of accuracy, the IONet framework takes approximately 3.3 times the cost of training compared to the $\phi-$DeepONet frameworks.

\begin{table}[h!]
\centering
\begin{tabular}{l c c c c c c c c c c}
\toprule
& \textbf{SE} 
& \textbf{CE} 
& \multicolumn{5}{c}{\textbf{Non-linear CE}} 
& \textbf{DeepONet}
& \textbf{IONet} \\
\cmidrule(lr){4-8}
&  &  & $D=1$ & $D=2$ & $D=3$ & $D=4$ & $D=5$ &  \\
\midrule
\textbf{Rel.~$L_2$ error}
& $7.11\text{e-}1$
& $8.25\text{e-}1$
& $7.13\text{e-}1$
& $4.32\text{e-}1$
& $8.52\text{e-}2$
& $6.29\text{e-}2$
& $5.58\text{e-}2$
& $6.63\text{e}0$ 
& $2.36\text{e-}2$ \\
\textbf{Cost}
& 0.89
& 1.00
& 1.01
& 1.12
& 1.13
& 1.21
& 1.30
& 1.36
& 3.38 \\
\bottomrule
\end{tabular}
\caption{Relative $L_2$ errors on the test set for the various $\phi$-DeepONet frameworks (along with the standard physics-informed DeepONet and IONet models) for the 1D problem with four interfaces and discontinuous input functions (Section~\ref{sec:1d_fourint_discoituous}).}\label{tab:1d_fourint_discont_errors}
\end{table}

\begin{figure}[!hbt]
    \centering
    \begin{subfigure}[b]{0.77\textwidth}
        \centering
        \includegraphics[width=\textwidth]{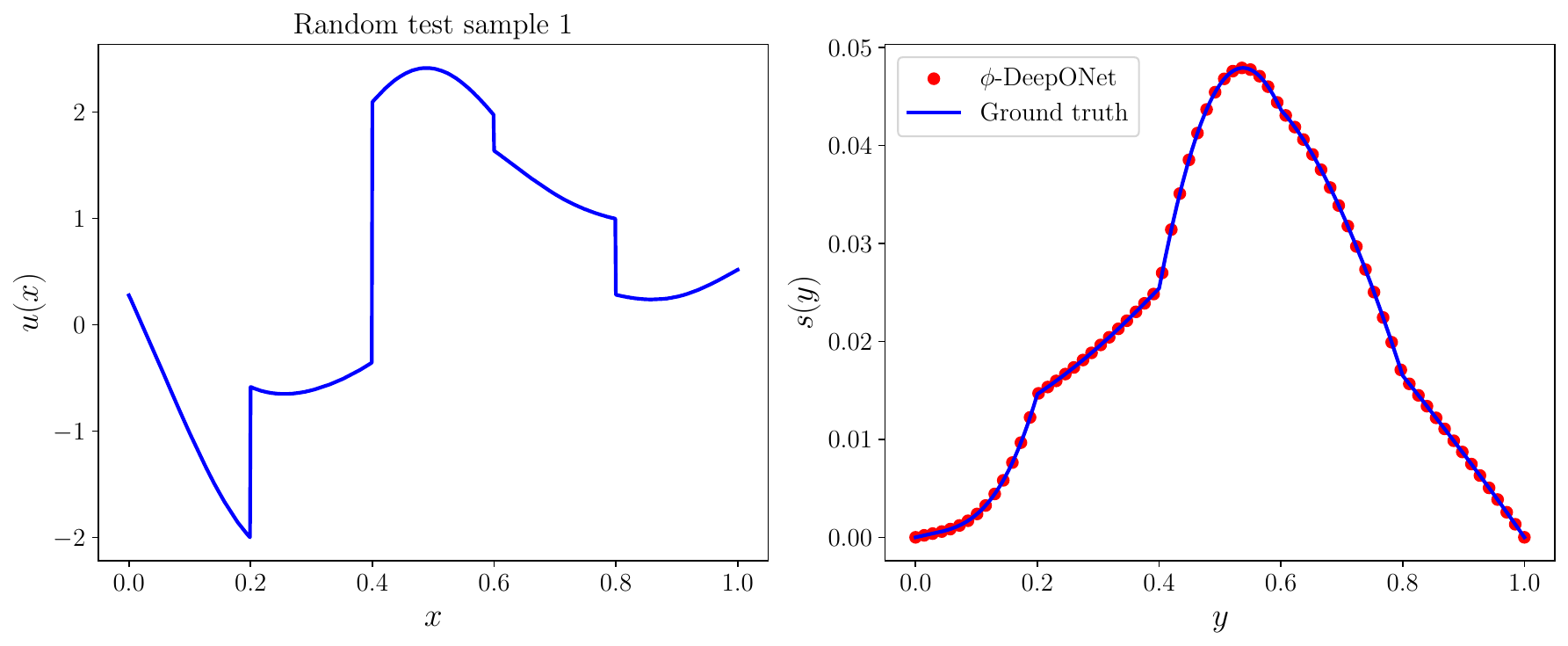}
        \caption{}
        \label{fig:1d_fourint_discontu_compare_test1}
    \end{subfigure}\\
    \centering
    \begin{subfigure}[b]{0.77\textwidth}
        \centering
        \includegraphics[width=\textwidth]{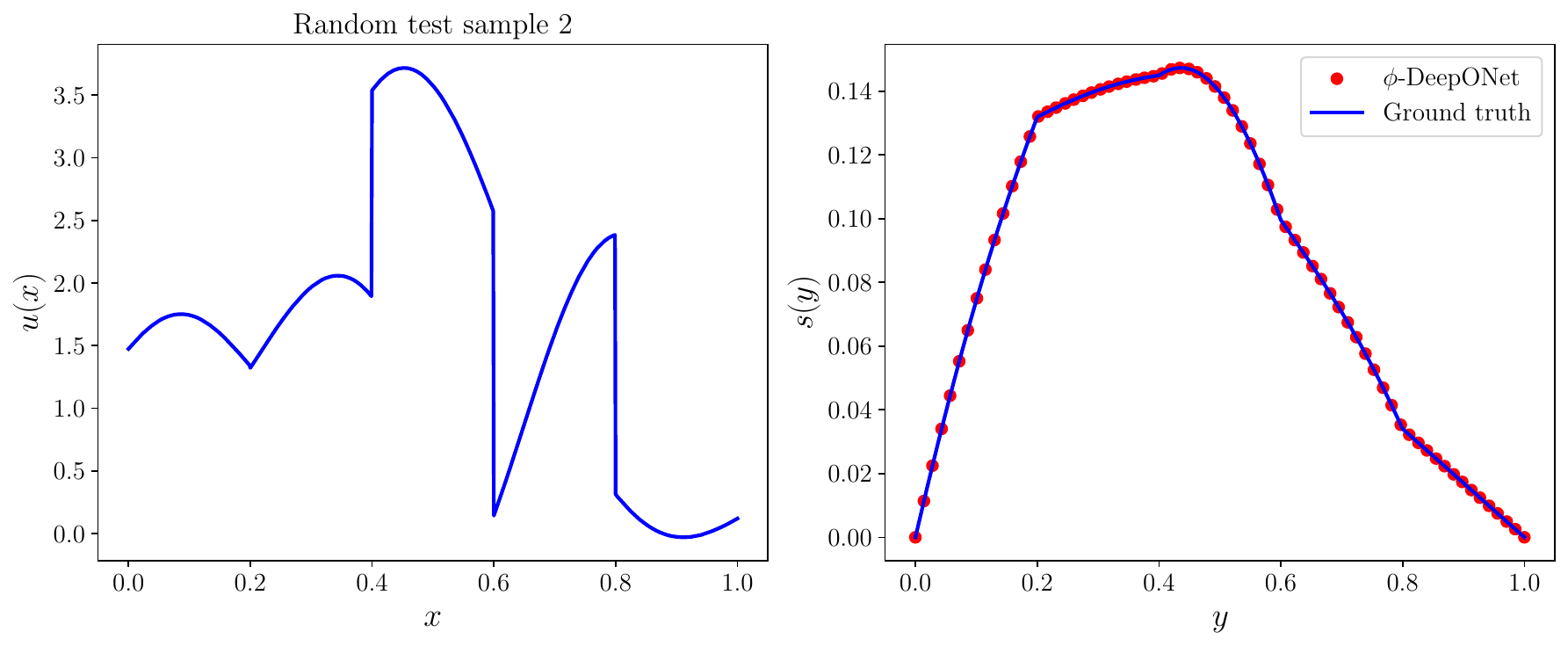}
        \caption{}
        \label{fig:1d_fourint_discontu_compare_test2}
    \end{subfigure}\\
    \centering
    \begin{subfigure}[b]{0.4\textwidth}
        \centering
        \includegraphics[width=\textwidth]{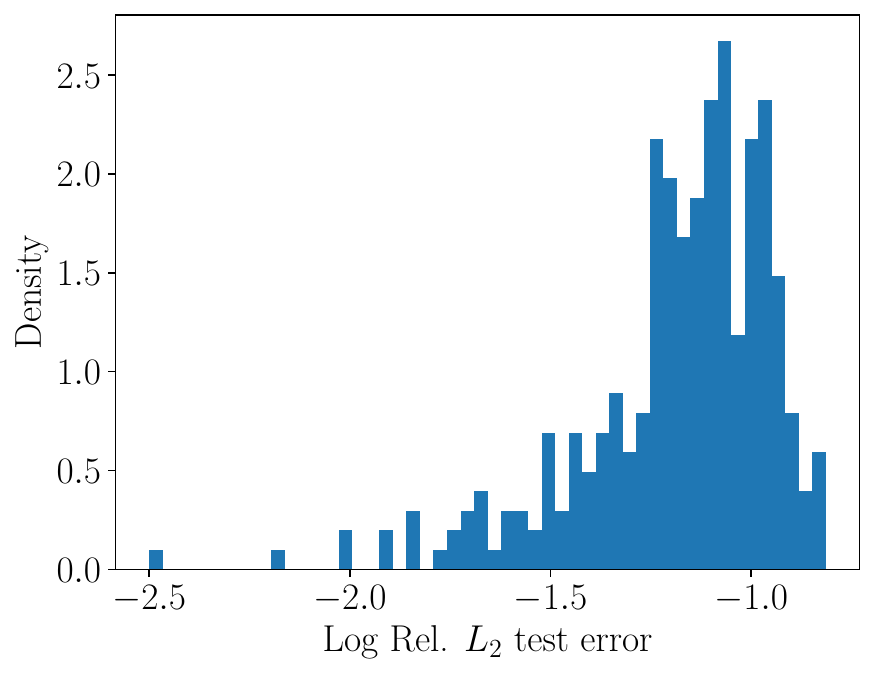}
        \caption{}
        \label{fig:1dfourint_discontu_errorplot}
    \end{subfigure}\\
    \caption{1D multiple-interface problem with discontinuous $u(x)$: performance of the $\phi$-DeepONet framework on two random test samples (subfigures (a) and (b)), along with the distribution of the test errors (subfigure (c)).}
    \label{fig:1d_fourint_discontu_examples_with_convergence}
\end{figure}

Figure~\ref{fig:1d_fourint_discontu_examples_with_convergence} shows the prediction of the trained $\phi-$DeepONet (non-linear CE with $D=3$) when given two random input functions from the test set. It is observed that the trained model produces approximations which are in good agreement with the ground truth (reference) solution which is obtained from a finite difference method. Figure~\ref{fig:1dfourint_discontu_errorplot} shows the distribution of the test errors.

\subsection{One-dimensional problem (with discontinuous outputs)}\label{sec:1d_oneint_disout}

We now reconsider the example from Section~\ref{sec:1d_oneint}, but in this case the solution field is also discontinuous at the interface. Specifically, we consider the one-dimensional domain $\Omega = [0,1]$ with an interface located at $\Gamma = 0.5$, which partitions $\Omega$ into two subdomains: $\Omega_1 = (0,\xi)$ and $\Omega_2 = (\xi,1)$. We seek to train the neural operator to approximate a piecewise-defined scalar field
\[
s(y) =
\begin{cases}
s_1(y), & y \in \Omega_1, \\
s_2(y), & y \in \Omega_2,
\end{cases}
\]
that satisfies a second-order PDE on each subdomain $\Omega_q$, given by
\begin{align}
  \begin{split}
      \frac{d}{dy}\left({\kappa_q} \frac{d{s_q}}{dy}\right)&= u_q \quad \text{in} ~ \Omega_q, \\
       s_{1} &= 0  \quad \text{at}~y=0, \\
       s_{2} &= 0  \quad \text{at}~y=1, \\
       \left\llbracket \frac{s}{H} \right\rrbracket &=0, ~\text{at}~ y={0.5}, \\
       \left \llbracket {\kappa\frac{ds}{dy}}  \right \rrbracket &=0 ~\text{at}~ y={0.5}. 
  \end{split}
\end{align}\label{eq:1d_oneint_dis}

The source terms $u_1(y)$ and $u_2(y)$ are modeled as independent realizations from a common Gaussian process prior. Although both source terms are drawn from the same prior, they are sampled independently and are therefore generally discontinuous across the interface. Here, $H_q > 0$ is a material-dependent coefficient, and $[\![\cdot]\!]$ denotes the jump across the interface, as usual. The material parameters used in this problem are $\kappa_1=2.0$, $\kappa_2=1.0$, $H_1=1.0$, and $H_2=2.0$.

\textbf{Physical interpretation:} This problem exhibits several important features. First, the primary variable $s(y)$ is generally discontinuous at the interface when $H_1 \neq H_2$. However, the transformed quantity $s/H$ remains continuous, and the diffusive flux $\kappa \, ds/dy$ is conserved across the interface. The stochastic forcing induces variability in the solution space, making this problem a suitable benchmark for evaluating operator learning methods in the presence of discontinuities in both the input and output fields. Physically, this problem corresponds to steady-state diffusion of a chemical species, where $s$ is the concentration and is related to the partial pressure $p$ through $s_i = H_i p$ \cite{wadiak1986application,castonguay2023modeling}. The interface conditions therefore correspond to continuity of partial pressure and continuity of flux.

We train the $\phi$-DeepONet variants to map the discontinuous inputs $u(x)$ to the outputs $s(y)$ using $N_\text{train}=5000$ training samples and evaluate them on $N_\text{test}=500$ test samples. Table~\ref{tab:1d_oneint_disout} reports the approximation accuracy in terms of the relative $L_2$ error on the test dataset. We observe that the SE and CE models achieve similar levels of accuracy, with errors on the order of $\mathcal{O}(10^{-2})$, whereas the non-linear models attain errors on the order of $\mathcal{O}(10^{-3})$. Among the non-linear CE models, we do not observe any substantial differences in accuracy. This may be because the present problem contains only a single interface, and therefore only two subdomains. As a result, choosing a latent dimension $D \geq 2$ primarily increases the representation dimension without providing a significant additional benefit. In terms of computational cost, all models exhibit broadly similar training costs.

\begin{figure}[!hbt]
    \centering
    \begin{subfigure}[b]{0.77\textwidth}
        \centering
        \includegraphics[width=\textwidth]{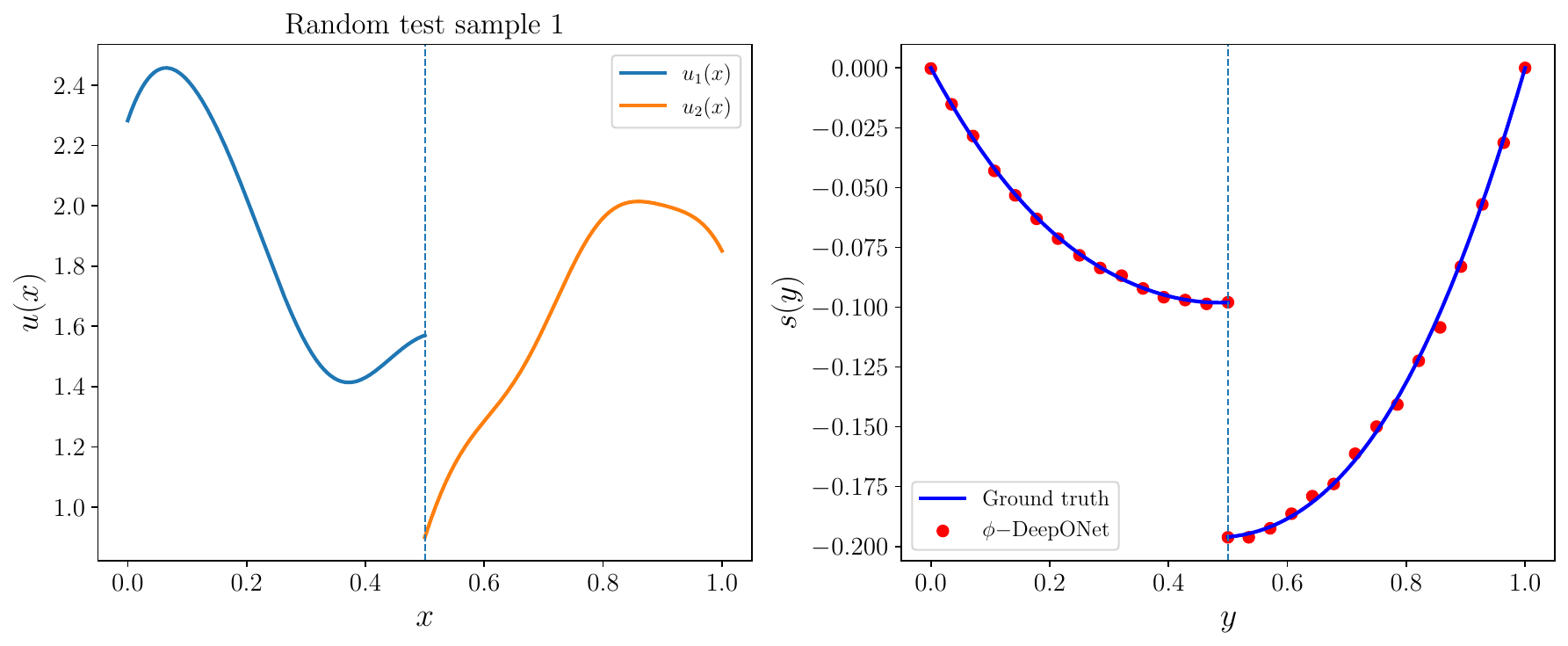}
        \caption{}
        \label{fig:1d_discontout_compare_test1}
    \end{subfigure}\\
    \centering
    \begin{subfigure}[b]{0.77\textwidth}
        \centering
        \includegraphics[width=\textwidth]{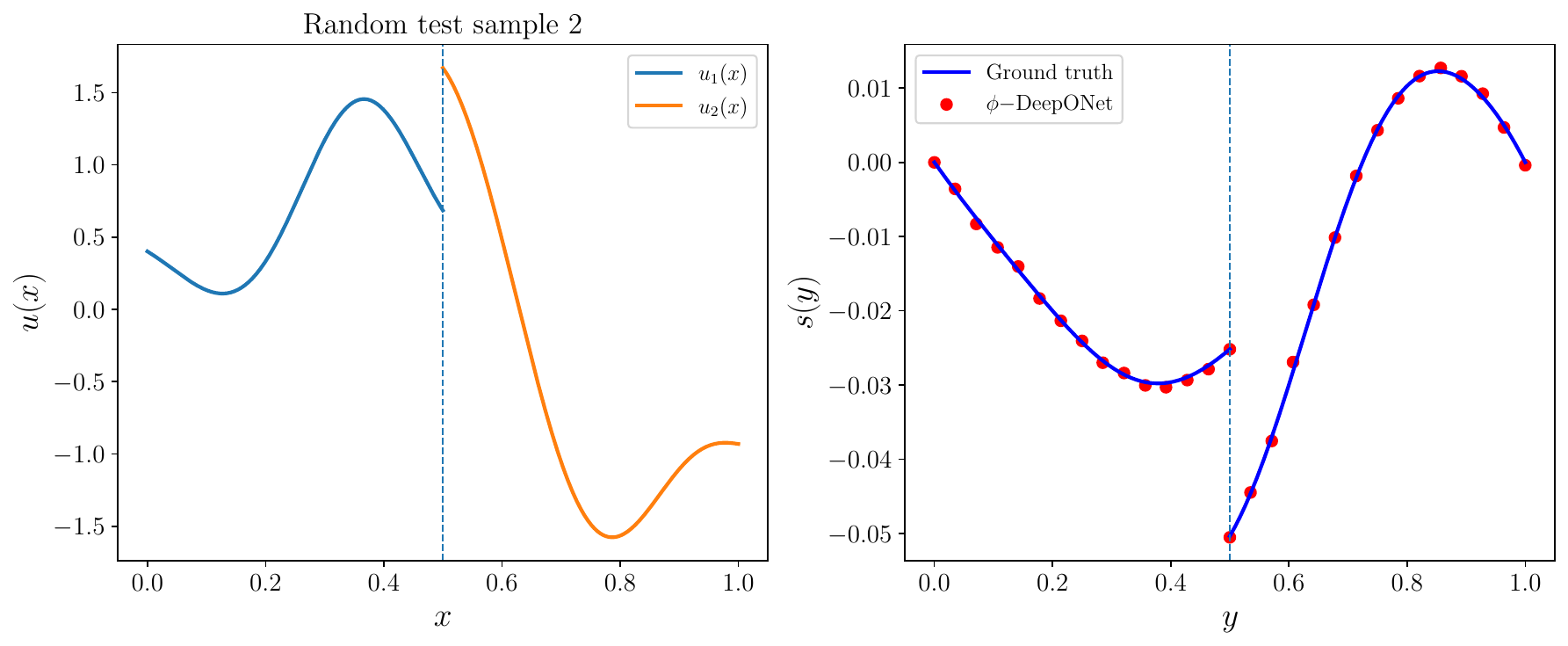}
        \caption{}
        \label{fig:1d_discontout_compare_test2}
    \end{subfigure}\\
    \centering
    \begin{subfigure}[b]{0.5\textwidth}
        \centering
        \includegraphics[width=\textwidth]{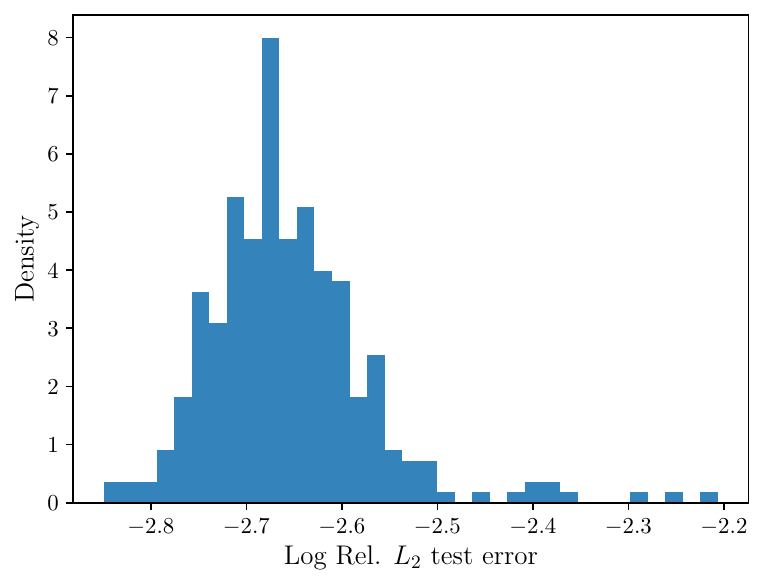}
        \caption{}
        \label{fig:1d_discontout_errorplot}
    \end{subfigure}\\
    \caption{1D problem with discontinuous inputs $u(x)$ and outputs $s(y)$: performance of the $\phi$-DeepONet framework on two random test samples (subfigures (a) and (b)), along with the distribution of the test errors (subfigure (c)).}
    \label{fig:1d_discontout}
\end{figure}

\begin{table}[h!]
\centering
\begin{tabular}{l c c c c c c c}
\toprule
& \textbf{SE} 
& \textbf{CE} 
& \multicolumn{5}{c}{\textbf{Non-linear CE}}  \\
\cmidrule(lr){4-8}
&  &  & $D=1$ & $D=2$ & $D=3$ & $D=4$ & $D=5$  \\
\midrule
\textbf{Rel.~$L_2$ error}
& $3.76\text{e-}2$
& $1.69\text{e-}2$
& $7.05\text{e-}3$
& $6.36\text{e-}3$
& $6.02\text{e-}3$
& $5.55\text{e-}3$
& $4.93\text{e-}3$ \\
\textbf{Cost}
& 0.95
& 1.00
& 1.04
& 1.11
& 1.07
& 1.12
& 1.14 \\
\bottomrule
\end{tabular}
\caption{Relative $L_2$ errors on the test set for the various $\phi$-DeepONet frameworks for the 1D problem with and discontinuous input and output functions (Section~\ref{sec:1d_oneint_disout}).}\label{tab:1d_oneint_disout}
\end{table}

\subsection{Two-dimensional problem with petal shaped interface}\label{sec:2d_petal}

\begin{figure}[!hbt]
    \centering
    \begin{subfigure}[b]{0.3\textwidth}
        \centering
        \includegraphics[width=\textwidth]{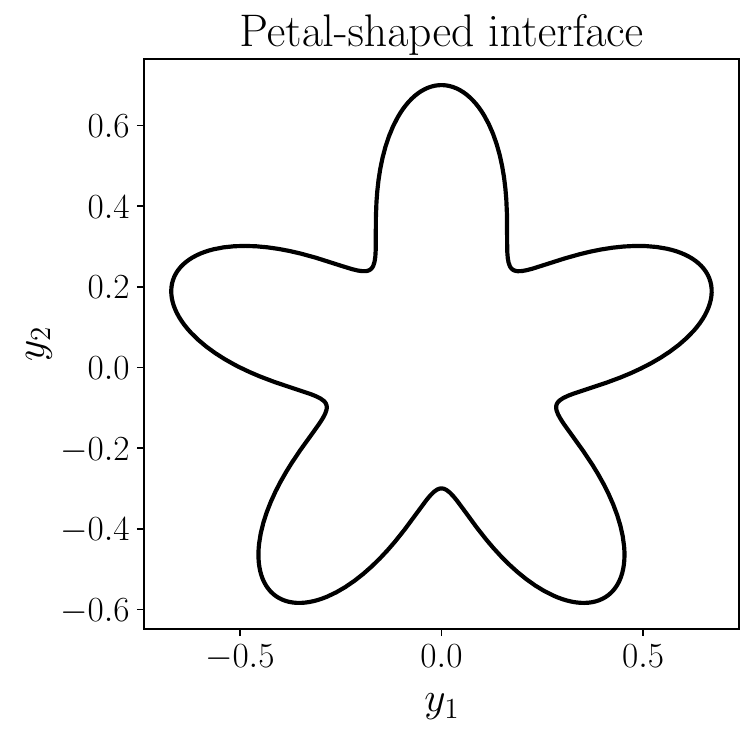}
        \caption{}
        \label{fig:2d_petal_shaped_interface}
    \end{subfigure}\\
    \begin{subfigure}[b]{0.32\textwidth}
        \centering
        \includegraphics[width=\textwidth]{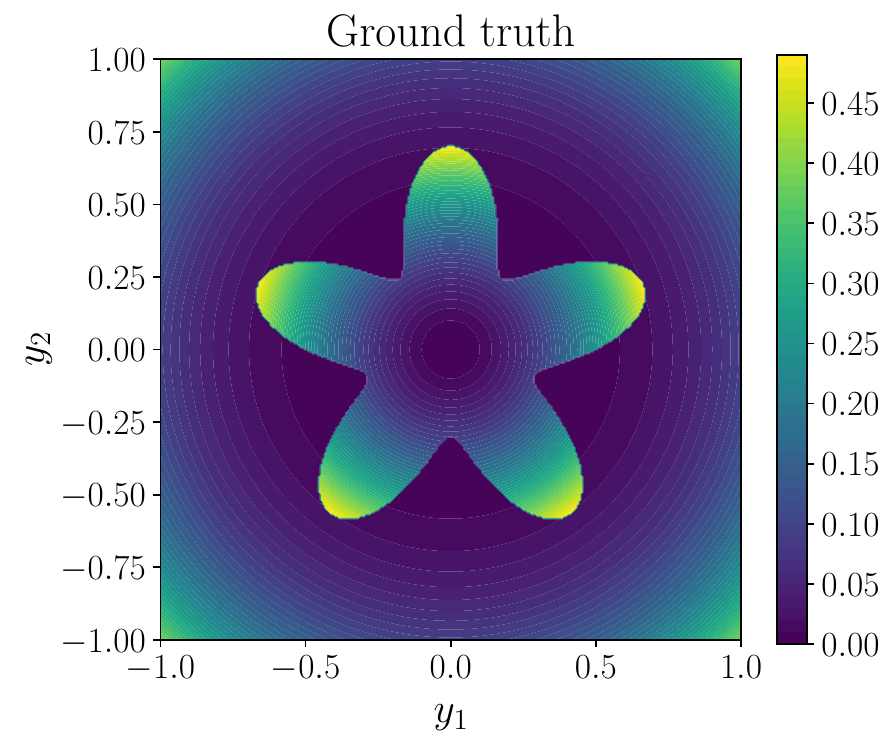}
        \caption{}
        \label{fig:2d_petal_shaped_ground}
    \end{subfigure}
    \hspace{2.0pt}
    \begin{subfigure}[b]{0.32\textwidth}
        \centering
        \includegraphics[width=\textwidth]{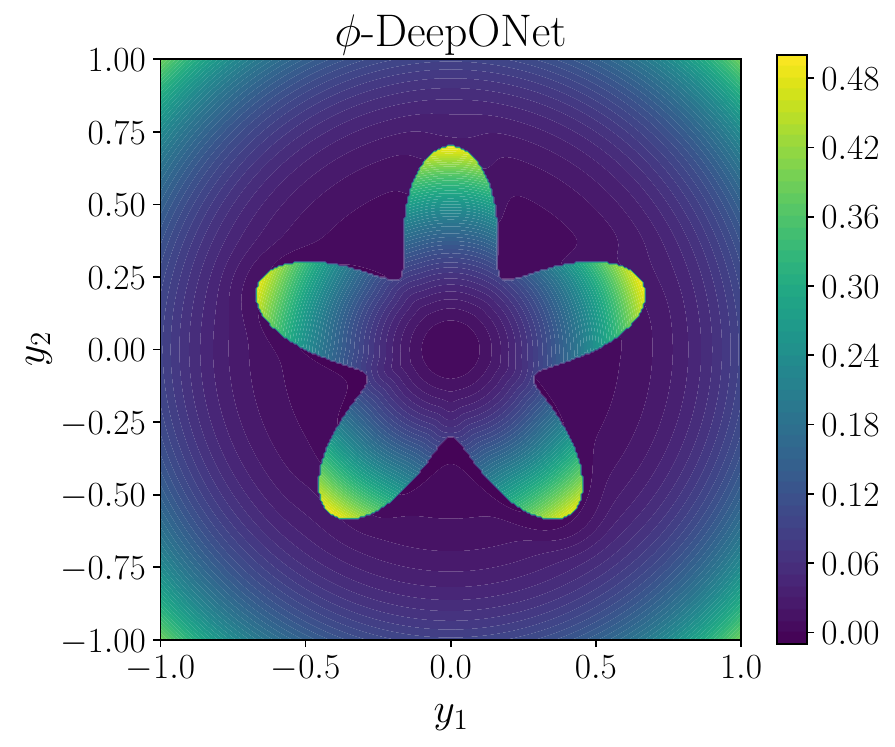}
        \caption{}
        \label{fig:2d_petal_shaped_phidon}
    \end{subfigure} 
    \hspace{2.0pt}
    \begin{subfigure}[b]{0.32\textwidth}
        \centering
        \includegraphics[width=\textwidth]{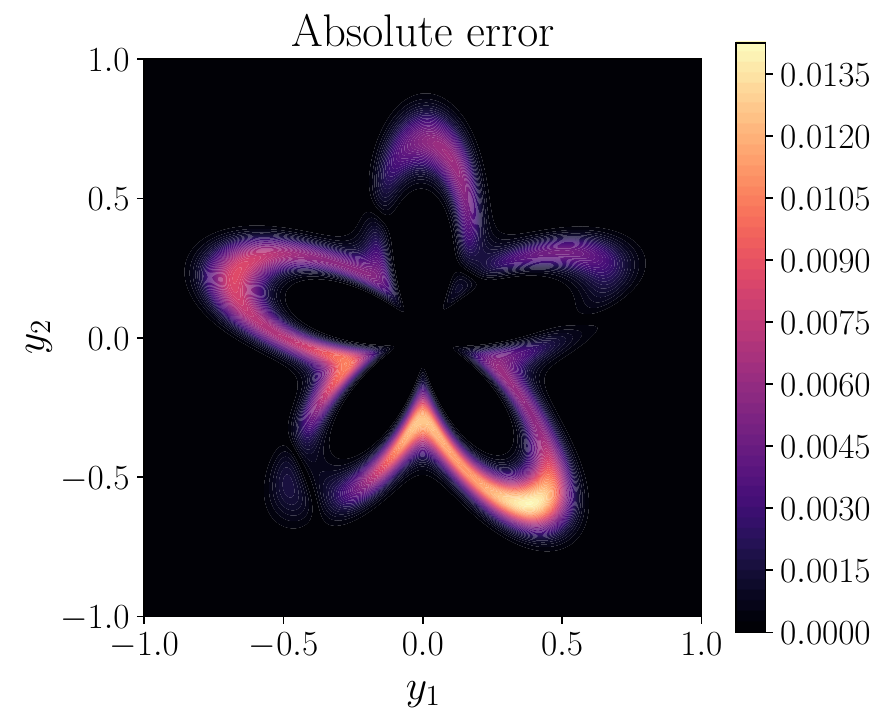}
        \caption{}
        \label{fig:2d_petal_shaped_error}
    \end{subfigure}\\
    \caption{2D petal-shaped interface: the predicition of the $\phi-$DeepONet framework compared to the ground truth along with the absolute error.}\label{fig:2d_petal_shaped}
\end{figure}

Finally, we consider a two-dimensional elliptic interface problem defined on the computational domain: $\Omega = [-1,1]\times[-1,1]$, with a petal-shaped interface $\Gamma_{\text{int}}$ whose geometry is described parametrically as
\begin{align}
  \begin{split}
    y_1(\theta) &= 0.02\sqrt{5}+(0.5+0.2\sin{5\theta})\cos{\theta},\\
    y_2(\theta) &= 0.02\sqrt{5}+(0.5+0.2\sin{5\theta})\sin{\theta},
  \end{split}
\end{align}
where $-\pi \leq \theta \leq \pi$. The domain is partitioned into two subdomains $\Omega_1$ and $\Omega_2$, corresponding to the interior and exterior of the interface, respectively (see Figure~\ref{fig:2d_petal_shaped_interface}). In the operator learning setting, we seek to learn a nonlinear operator $\mathcal{G}: u(\boldsymbol{x}) \mapsto s(\boldsymbol{y})$ which maps an input forcing function $u(\boldsymbol{x})$, defined over coordinates $\boldsymbol{x} = (x_1,x_2) \in \Omega$, to a solution field $s(\boldsymbol{y})$, evaluated at coordinates $\boldsymbol{y} = (y_1,y_2) \in \Omega$. The solution satisfies the Poisson equation in each subdomain:
\begin{align}\label{eq:petal_shaped_pde}
    \begin{split}
        &\nabla_{\mathbf{y}}\cdot (\kappa_\text{m} \nabla_{\mathbf{y}} s_\text{m})= u_\text{m}(\mathbf{y}) \qquad \text{in}~~\Omega_\text{m},\\
        &s_2 = 0.1(y_1^2+y_2^2)^2 - 0.01\log\left(2\sqrt{y_1^2+y_2^2}\right) \quad \text{on}~~\partial\Omega_2^\text{d},\\
        &\llbracket s \rrbracket = 0.1(y_1^2+y_2^2)^2 -0.01\log\left(\sqrt{y_1^2+y_2^2}\right)-(y_1^2+y_2^2), \qquad \text{on}~\Gamma_\text{int},\\
        &\llbracket \kappa \nabla_{\mathbf{y}} s \rrbracket \cdot \mathbf{n} = \left(4(y_1^2+y_2^2) - 0.1(y_1^2+y_2^2)^{-1}-2\right)(y_1 n_{1}+y_2 n_{2}) \qquad \text{on}~\Gamma_\text{int},
    \end{split}
\end{align}
where $\partial\Omega_2^\text{d}$ denotes the external boundary of the domain with diffusion coefficients $\kappa_1=4$ and $\kappa_2=10$. The forcing function is defined piecewise over the subdomains as
\[
u(\mathbf{x}; p_1,p_2) =
\begin{cases}
p_1, & \mathbf{x} \in \Omega_1,\\
p_2(x_1^2+x_2^2), & \mathbf{x} \in \Omega_2,
\end{cases}
\]
where $(p_1,p_2) \in \mathcal{P} = \{ 2, 20 \}$ are scalar parameters controlling the magnitude of the forcing in each subdomain. By varying $(p_1,p_2)$, we obtain a class of interface problems with distinct solutions. It is to be noted that, the vector $\boldsymbol{p}=[p_1,p_2]$ act as two-dimensional embedding of the input function $u(\boldsymbol{x})$ and instead of using two branch-nets, one could easily use a single branch-net with two inputs. However, we do not do so because in real-life cases we would not have access to this embedding and the input functions would generally be available to us with sensors places across the domain. Thus we sample the function across 40  sensors (20 in each sub-domain) to get the vector valued input function. 

For a particular choice of parameters $(p_1^\star,p_2^\star) = (4,16)$, the forcing reduces to
\[
u(\mathbf{x}; p_1^\star,p_2^\star) =
\begin{cases}
4, & \mathbf{x} \in \Omega_1,\\
16(x_1^2+x_2^2), & \mathbf{x} \in \Omega_2,
\end{cases}
\]
for which the problem admits the known analytical solution:
\begin{align}
  s(\mathbf{y}) =
    \begin{cases}
      y_1^2+y_2^2, & \mathbf{y} \in \Omega_{1}, \\
      0.1(y_1^2+y_2^2)^2 -0.01\log\left(2\sqrt{y_1^2+y_2^2}\right), & \mathbf{y} \in \Omega_{2}.
    \end{cases}
\end{align}
The operator $\mathcal{G}$ is trained on a collection of $N_{train}=80$ forcing functions generated by sampling $(p_1,p_2)$ from the predefined parameter space $\mathcal{P}$, excluding the reference pair $(p_1^\star,p_2^\star)$. The learned operator is then evaluated on the held-out parameter pair $(p_1^\star,p_2^\star)$, for which the analytical solution is available (thus $N_{test}=1$). This enables a direct assessment of the operator's predictive accuracy on an unseen PDE instance. Figure~\ref{fig:2d_petal_shaped} shows the prediction of the $\phi-$DeepONet model on this particular test-case, compared to the ground truth and the absolute error. From the plot we observe that the model is able to approximate the solution of this test PDE with sufficient accuracy, with errors only accumulating slightly at the interface. The relative $L_2$ error for this problem on the test example of $2.1e$-1. This is relatively higher than the previous case, because the interface geometry along with the jump conditions are more complex compared to the previous example. 

\begin{figure}[!hbt]
    \centering
    \begin{subfigure}[b]{0.32\textwidth}
        \centering
        \includegraphics[width=\textwidth]{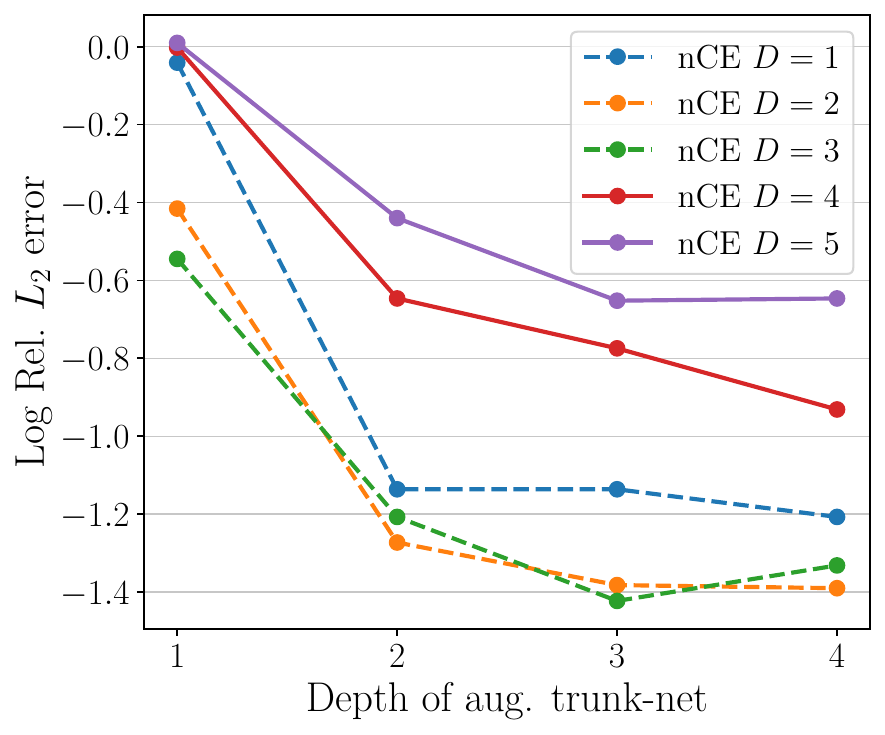}
        \caption{}
        \label{fig:petal_interface_trunkdepth}
    \end{subfigure}
    \hspace{2.0pt}
    \begin{subfigure}[b]{0.32\textwidth}
        \centering
        \includegraphics[width=\textwidth]{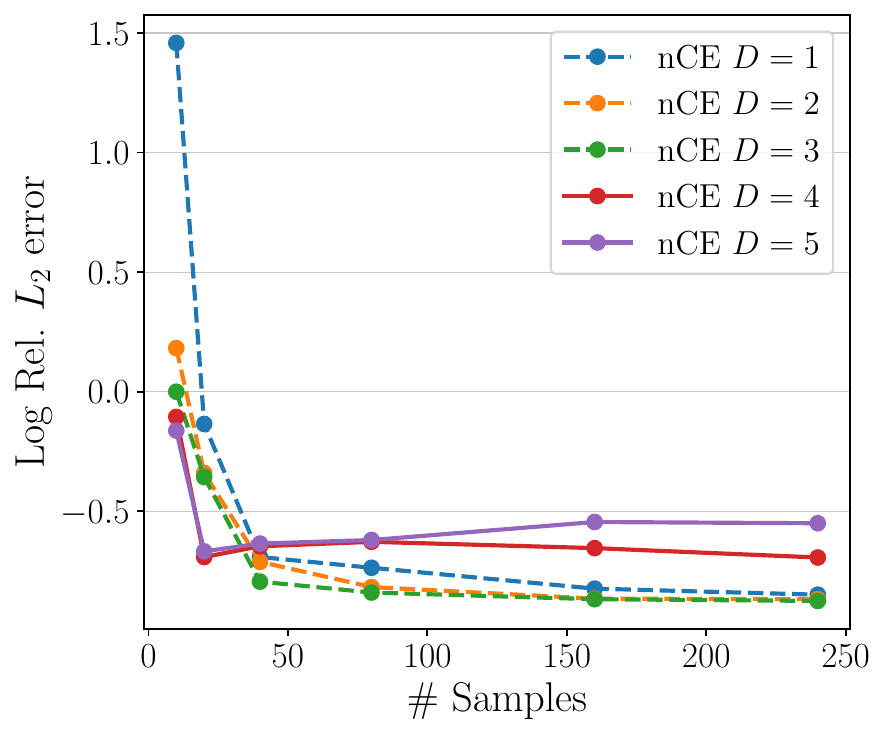}
        \caption{}
        \label{fig:petal_interface_samples}
    \end{subfigure} 
    \hspace{2.0pt}
    \begin{subfigure}[b]{0.32\textwidth}
        \centering
        \includegraphics[width=\textwidth]{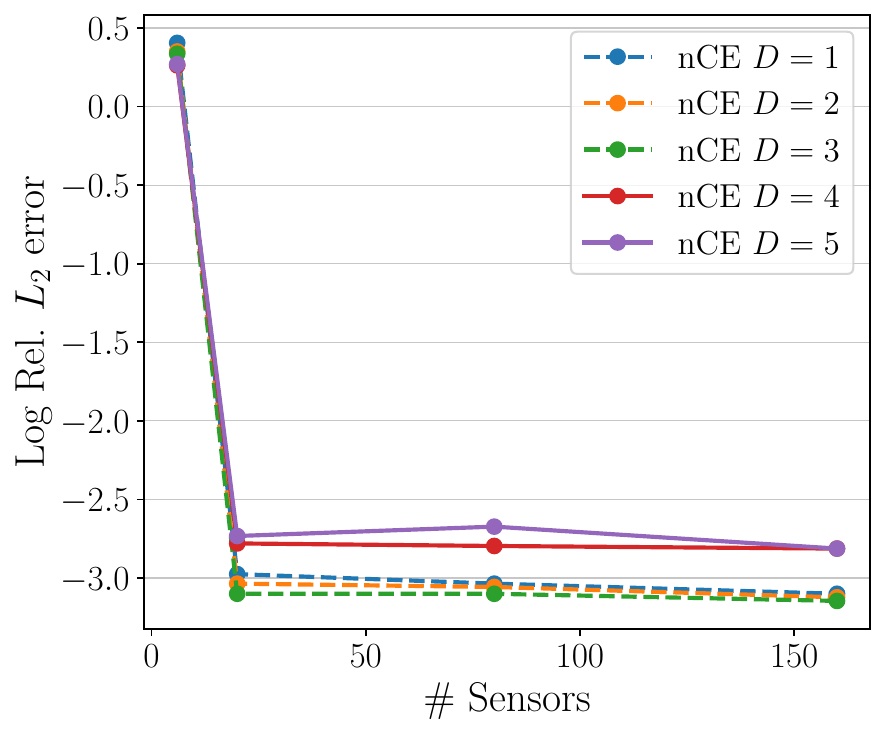}
        \caption{}
        \label{fig:petal_interface_sensors}
    \end{subfigure}\\
    \caption{2D petal-shaped interface: An ablation study showing the performance of various variants of $\phi$-DeepONet. (a) Relative $L_2$ error for various depths of the augmented trunk network, (b) relative $L_2$ error versus the number of training samples, and (c) relative $L_2$ error versus the total number of sensors at which the input function is sampled.}\label{fig:2d_petal_shaped_ablation}
\end{figure}

Table~\ref{tab:2d_petal} shows the relative $L_2$ errors across various variants of the $\phi$-DeepONet framework. We observe that, although the computational cost remains comparable across all models, the non-linear CE variants begin to exhibit increased error as the latent dimension exceeds $D \geq 3$. This behavior is likely due to over-parameterization in the latent space, particularly for problems involving only two subdomains, where higher-dimensional embeddings may introduce redundancy rather than additional expressive benefit. To further investigate this trend, we conduct the ablation study shown in Fig.\ref{fig:2d_petal_shaped_ablation}, examining the effects of trunk network depth, training data size, and input function sensor resolution across different embedding dimensions $D$. The results consistently support the observations from Table~\ref{tab:2d_petal}. Increasing the embedding dimension from $D=1$ to $D=3$ generally improves performance across all settings; however, further increasing the dimension to $D=4$ and $D=5$ leads to degraded accuracy, often performing worse than lower-dimensional counterparts. This effect is particularly pronounced in Fig.~\ref{fig:2d_petal_shaped_ablation}(a), where, despite the overall decrease in relative $L_2$ error with increasing trunk network depth, the error curves corresponding to $D=4$ and $D=5$ consistently lie above those for smaller embedding dimensions across all depths. This trend confirms that the benefits of increasing embedding dimension saturate at moderate values and that larger latent spaces can negatively impact performance in this setting. Overall, these findings suggest that a compact embedding (e.g., $D=2$ or $D=3$) is sufficient to capture the interface structure effectively while avoiding unnecessary model complexity.

\begin{table}[h!]
\centering
\begin{tabular}{l c c c c c c c }
\toprule
& \textbf{SE} 
& \textbf{CE} 
& \multicolumn{5}{c}{\textbf{Non-linear CE}}  \\
\cmidrule(lr){4-8}
&  &  & $D=1$ & $D=2$ & $D=3$ & $D=4$ & $D=5$  \\
\midrule
\textbf{Rel.~$L_2$ error}
& $7.25\text{e-}1$
& $5.83\text{e-}1$
& $3.26\text{e-}1$
& $3.12\text{e-}1$
& $2.24\text{e-}1$
& $5.91\text{e-}1$
& $7.44\text{e-}1$ \\
\textbf{Cost}
& 0.98
& 1.00
& 1.01
& 1.14
& 1.19
& 1.20
& 1.20 \\
\bottomrule
\end{tabular}
\caption{Relative $L_2$ errors on the test example for the various $\phi$-DeepONet frameworks for the 2D problem with a petal-shaped interface (Section~\ref{sec:2d_petal}).}\label{tab:2d_petal}
\end{table}

\section{Conclusion}\label{sec:conclusions}

In this work, we introduce a physics-informed neural operator, termed $\phi$-DeepONet, for learning mappings from continuous or discontinuous input functions to non-smooth output fields. Such problems arise naturally in interface settings that are ubiquitous in science and engineering, where solution fields exhibit strong or weak discontinuities due to piecewise-constant material properties and spatially varying, discontinuous forcing functions. Unlike traditional physics-informed learning approaches that rely on explicit domain decomposition, the proposed framework implicitly incorporates domain structure by learning a latent field $\boldsymbol{\phi}$, which is provided as an additional input to an augmented trunk network. This trunk network is coupled with multiple branch networks, each responsible for encoding input functions associated with different subdomains. The governing physics and interface continuity conditions are enforced through soft constraints in the loss function, enabling the model to learn solutions that respect both the PDE and the interface behavior. Numerical results demonstrate that, while standard physics-informed DeepONet struggles to accurately capture discontinuities, $\phi$-DeepONet achieves significantly improved performance, reducing test errors by up to two orders of magnitude. Furthermore, when compared to IONet, a domain-decomposition-based DeepONet variant, the proposed approach attains comparable or superior accuracy with substantially lower computational cost (in the range of 1.5-4 times depending on the complexity of the problem). Overall, these results highlight the effectiveness of embedding interface information directly within the operator learning framework, enabling accurate and efficient modeling of complex discontinuous systems without the need for explicit domain partitioning.

Despite its strong performance, the proposed framework has several limitations that motivate future work. First, the method relies on prior knowledge of subdomain partitions to construct the categorical embedding, which may not be available in problems with unknown or evolving interfaces. Second, the latent representation is piecewise constant within each subdomain, which may limit its ability to capture finer-scale variations near interfaces or within heterogeneous regions. Additionally, the choice of embedding dimension $D$ remains problem-dependent, with larger values potentially leading to over-parameterization without corresponding performance gains. Furthermore, interface conditions are enforced in a soft manner through the loss function, which may not guarantee strict satisfaction in all cases. Future work will focus on developing more flexible and data-driven representations of interface structure, including learning continuous latent fields or implicit interface representations that remove the need for explicit domain partitioning. Extensions to stochastic and uncertain interface problems, as well as more complex multiphysics systems, are also promising directions. Finally, improving the theoretical understanding of the role of the latent embedding and its optimal dimensionality remains an important open question.

\section*{Acknowledgments}

Michael Shields acknowledges the support of the U.S. Department of Energy, Office of Science, Office of Advanced Scientific Computing Research, under Award No.~DE-SC0024162. Pratanu Roy and Stephen Castonguay's work was performed under the auspices of the U.S. Department of Energy by Lawrence Livermore National Laboratory under Contract DE-AC52–07NA27344 and was supported by the LLNL-LDRD program under project number 25-ERD-052. LLNL release number LLNL-JRNL-2017790.

\bibliographystyle{unsrtnat}   
\bibliography{references}

\end{document}